\documentclass[journal]{IEEEtran}
\ifCLASSINFOpdf
\else
\fi
\usepackage{cite}
\usepackage{amsmath,amssymb,amsfonts}
\usepackage{algorithm} 
\usepackage{graphicx}
\usepackage{caption}
\usepackage{subcaption}
\usepackage{textcomp}
\usepackage{array}
\usepackage{textcomp}
\usepackage{stfloats}
\usepackage{url}
\usepackage{verbatim}
\usepackage{amsthm, bm, hyperref, mathrsfs, indentfirst, longtable}
\usepackage{svg}
\usepackage{longtable}
\usepackage{tabularx}
\usepackage{makecell}
\usepackage{booktabs,float}
\usepackage{bbm}
\usepackage{balance}
\usepackage{threeparttable}
\usepackage{multirow}
\usepackage{booktabs}
\usepackage[table]{xcolor}
\usepackage{multirow}
\begin{document}
\title{JEPA-MSAC: A Joint-Embedding Predictive Architecture for Multimodal Sensing-Assisted Communications}

\author{
    Can Zheng, 
    Jiguang He,~\IEEEmembership{Senior Member,~IEEE,}
    Guofa Cai,~\IEEEmembership{Senior Member,~IEEE,}
    Nannan Li,\\
    Mehdi Bennis,~\IEEEmembership{Fellow,~IEEE,} 
    Henk Wymeersch,~\IEEEmembership{Fellow,~IEEE,} 
    M\'erouane Debbah,~\IEEEmembership{Fellow,~IEEE} 
    \thanks{Can Zheng and Jiguang He are with the School of Computing and Information Technology, Great Bay University, Dongguan Key Laboratory for Intelligence and Information Technology, and Great Bay Institute for Advanced Study (GBIAS), Dongguan 523000, China (e-mail: zc331\_@korea.ac.kr, jiguang.he@gbu.edu.cn).}
    \thanks{Guofa Cai is with the School of Information Engineering, Guangdong University of Technology, Guangzhou 510006, China (e-mail: caiguofa2006@gdut.edu.cn).}
    \thanks{Nannan Li is with the  School of Computer Science and Engineering, Macau University of Science and Technology, Macau 999078, China (e-mail: nnli@must.edu.mo).}
    \thanks{Mehdi Bennis is with the Faculty of Information Technology and Electrical Engineering, Centre for Wireless Communications, University of Oulu, 90570 Oulu, Finland (e-mail: mehdi.bennis@oulu.fi).}
    \thanks{Henk Wymeersch is with the Department of Electrical Engineering, Chalmers University of Technology, Gothenburg SE-412 96, Sweden. (emails: henkw@chalmers.se).}
    \thanks{M\'erouane Debbah is with  the Research Institute for Digital Future, Khalifa University, 127788 Abu Dhabi, UAE (email: merouane.debbah@ku.ac.ae)}
}

\maketitle
\begin{abstract}
    Future wireless systems increasingly require predictive and transferable representations that can support multiple physical-layer (PHY) tasks under dynamic environments. However, most existing supervised learning-based methods are designed for a single task, which leads to high adaptation cost. To address this issue, we propose a joint-embedding predictive architecture for multimodal sensing-assisted communications (JEPA-MSAC), a self-supervised multimodal predictive representation learning framework for wireless environments. The proposed framework first maps multimodal sensing and communication measurements into a unified token space, and then pretrains a shared backbone using temporal block-masked JEPA to learn a predictive latent space that captures environment dynamics and cross-modal dependencies. After pretraining, the backbone is frozen and reused as a general future-feature generator, on top of which lightweight task heads are trained for localization, beam prediction, and received signal strength indicator (RSSI) prediction. Extensive experiments show the latent state supports accurate multi-task prediction with low adaptation cost. Additionally, ablation studies reveal its scaling behavior and the impact of key pretraining setups.
\end{abstract}


\IEEEpeerreviewmaketitle

\section{Introduction}
    
    % 物理层、6G -> AI

    \IEEEPARstart{A}{s} wireless systems evolve toward sixth-generation (6G), integrating artificial intelligence (AI) into physical-layer (PHY) communications has emerged as a crucial shift to overcome the limitations of traditional model-based signal processing. This trend is reflected in recent industrial and standardization efforts, including 3rd generation partnership project (3GPP) studies on AI for the new radio (NR) air interface and the AI-radio access network (RAN) initiative \cite{3GPPTR38843,AIRAN,AIRAN2,AIRAN3}.
    % 物理层 Why AI? why LLMs?
    Data-driven deep learning (DL) has already shown great success in PHY-tasks by learning directly from complex environments \cite{DLPHY,CsiNet,ComNet,DLCE}. Building on this, the focus is rapidly shifting toward large AI models (LAMs) to achieve stronger generalization and flexible multimodal processing \cite{lam4phy,LLM4CP,BPLLM,BeamLLM,m2beamllm}. Recent advancements further push beyond adapting general-purpose LAMs, exploring wireless-native foundation models pretrained directly on wireless data \cite{WirelessGPT,lwm,LWLM}. Consequently, the PHY design paradigm is transitioning from isolated, single-task models toward unified, shared backbones capable of supporting multiple tasks \cite{llm4wm,llm4mt}.

    % 感知辅助通信
    Meanwhile, sensing-assisted communications have become increasingly important. In dynamic environments, PHY behaviors are shaped not only by radio frequency (RF) signals but also by external factors like user geometry, object motion, and blockage evolution \cite{deepsense6g}. These physical dynamics critically impact tasks such as user localization, beam prediction, and link-quality forecasting. Consequently, out-of-band (OOB) modalities, including cameras, radar, LiDAR, and positioning signals, provide essential geometric and environmental context that in-band RF measurements struggle to capture \cite{vision,radar,Lidar_,position,sub6ghz}. Processing such diverse data makes DL particularly effective for discovering cross-modal correlations and enhancing communication performance \cite{multimodal,moe}.

    % 感知辅助通信有什么问题：多模态融合只为了解决某个特定问题，泛化性差
    However, in many existing sensing-aided communication methods, multimodal fusion is primarily designed for a specific downstream objective, with multimodal inputs combined to improve a single target task. While effective for individual tasks, such designs often fail to capture the shared scene dynamics behind the observations. Recent LAM-based approaches alleviate this limitation only partially, as their main advantage still lies in stronger representation learning and task adaptation rather than explicit modeling of how the physical environment evolves over time \cite{LLMWM, largeactionmodel, embodiedai}. This suggests that instead of merely building a stronger task interface or a larger shared backbone, a more suitable direction is to learn a predictive latent representation of the underlying scene dynamics. To turn this intuition into a scalable framework, a world model-centric formulation provides a natural choice. By learning the latent state of the environment and its temporal evolution, future communication-relevant quantities, including location, beam decisions, and signal power, can be inferred from latent scene dynamics rather than direct observation-to-label mapping. This is particularly suitable for sensing-assisted wireless systems, where motion, geometry, blockage, and multimodal context jointly shape communication behavior \cite{World_Model,World_Model2}.

    Based on joint embedding prediction architectures (JEPA) \cite{I_JEPA,V-JEPA,V_JEPA2}, we propose \textbf{a JEPA for multimodal sensing-assisted communications (JEPA-MSAC)}, which learns a shared predictive latent state from multimodal observations and reuses it for downstream PHY-tasks. Rather than reconstructing raw inputs, it models temporally consistent scene dynamics with a frozen backbone and lightweight task heads.
    The main contributions of this work are summarized as follows:
    %\begin{itemize}
    %    \item We propose JEPA-MSAC, a JEPA-based world model for sensing-assisted wireless communications, which formulates sensing-assisted PHY-tasks as shared predictive state learning rather than task-specific multimodal fusion. We develop a unified multimodal representation framework that maps vision, radar, LiDAR, GPS, and RF observations into a homogeneous latent token space via {\color{blue}modality}-specific tokenizers, and pretrain the backbone to capture communication-relevant environment dynamics in latent space.
        
    %    \item On top of the pretrained world model, we develop a frozen-backbone and lightweight-head adaptation framework for multiple PHY-tasks, where the same predictive latent state is reused for localization, beam prediction, and received signal strength indicator (RSSI) prediction. We further introduce a geometry-guided cascaded design that injects localization-aware spatial priors into beam prediction and RSSI prediction.
        
    %    \item Extensive experiments demonstrate that the learned predictive latent state supports accurate and stable multi-task prediction with low adaptation cost. In addition, ablation studies reveal the scaling behavior of the latent representation and clarify the impact of key pretraining and temporal-context configurations, such as the masking ratio and historical observation window, on PHY-task performance.
    %\end{itemize}
    \begin{itemize}
        \item \textbf{Communication-oriented predictive latent state.} We propose JEPA-MSAC, a JEPA-based predictive representation learning framework for sensing-assisted wireless communications. Our main novelty is to learn a shared predictive latent representation via temporal block-masked JEPA. Instead of task-specific fusion or raw data reconstruction, we map multimodal inputs into a unified token space.
        
        \item \textbf{Frozen backbone and efficient task-specific heads.} On top of the pretrained backbone, we develop an efficient adaptation framework. We reuse a single frozen backbone across multiple PHY-tasks. The same predictive latent state supports localization, beam prediction, and received signal strength indicator (RSSI) prediction using lightweight task heads.
        
        \item \textbf{Localization-guided cascading.} We further introduce a geometry-guided cascaded design. To improve communication-related prediction, this design injects the predicted locations from the localization head into the beam and RSSI prediction heads.
        
        \item Extensive experiments demonstrate that the learned predictive latent state supports accurate and stable multi-task prediction with low adaptation cost. In addition, ablation studies reveal the scaling behavior of the latent representation and clarify the impact of key pretraining configurations, such as the masking ratio and pattern, on PHY-task performance.
    \end{itemize}

    %The remainder of this paper is organized as follows. Section~\ref{sec:system_model} introduces the system model and problem formulation. Section~\ref{sec:architecture} details the proposed JEPA-MSAC framework. Section~\ref{sec:simulation} provides simulation results. Finally, Section~\ref{sec:conclusion} concludes the paper.

    \textit{Notations}: Bold lowercase letters denote vectors (e.g., $\mathbf{x}$), and bold uppercase letters denote matrices or tensors (e.g., $\mathbf{X}$). Calligraphic letters denote sets (e.g., $\mathcal{X}$). The symbols $(\cdot)^\mathsf{T}$ and $(\cdot)^\mathsf{H}$ denote the transpose and Hermitian transpose, respectively. The sets of real and complex numbers are denoted by $\mathbb{R}$ and $\mathbb{C}$, respectively. The notations $||\cdot||_{1}$ and $||\cdot||_{2}$ denote the $\ell_{1}$ norm and Euclidean ($\ell_{2}$) norm, while $|\cdot|$ represents the absolute value or magnitude. The operator $\arg\max$ returns the maximum of a function. The operators $\ominus$ and $\oslash$ denote element-wise subtraction and element-wise division, respectively. The notation $[\cdot;\cdot]$ denotes representation concatenation. $\mathbb{E}[\cdot]$ denotes the mathematical expectation, and $\mathbb{P}(\cdot)$ represents probability. The indicator function is represented by $\mathbbm{1}_{\{\cdot\}}$, and $\lfloor\cdot\rfloor$ denotes the floor operation.
    
\section{System Model}
\label{sec:system_model}

    \begin{figure}
        \centering
        \captionsetup{font=footnotesize}
        \includegraphics[width=0.5\textwidth]{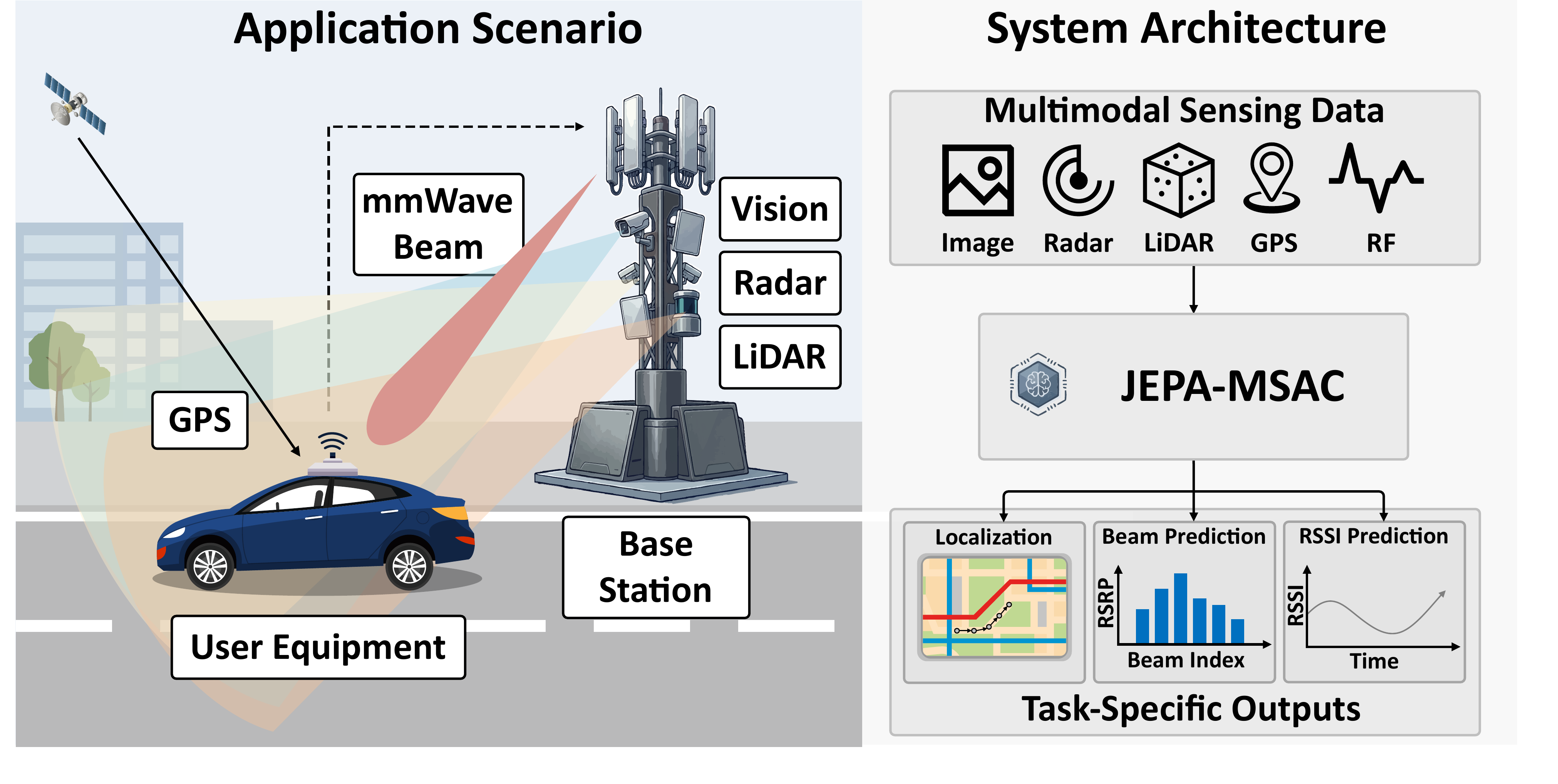}
        \caption{System architecture of sensing-assisted mmWave V2I communications. The JEPA-MSAC processes multimodal sensing and communication data to perform specific PHY-tasks, including localization, beam prediction, and RSSI prediction.}
    \label{fig:scenario}
    \end{figure}

    In this section, we present the system model for the proposed sensing-assisted millimeter-wave (mmWave) vehicle-to-infrastructure (V2I) communication scenario, as shown in Fig.~\ref{fig:scenario}. The framework encompasses the directional beamforming architecture, the localization model, and the multi-modal sensory observation space.
    
    \subsection{mmWave V2I Communication and ULA Beamforming}
    Consider a typical V2I downlink communication system where a base station (BS) serves a mobile user equipment (UE) vehicle. The BS is equipped with a uniform linear array (ULA) consisting of $N_\mathrm{Ant}$ antenna elements, while the UE is assumed to  operate in an omnidirectional receive mode to focus on the BS's transmit beamforming alignment.
    
    The BS employs an analog beamforming architecture relying on a predefined discrete codebook $\mathcal{W} = \{\mathbf{w}_1, \mathbf{w}_2, \dots, \mathbf{w}_K\}$, where $K$ denotes the total number of candidate beams. For a ULA, the steering vector $\mathbf{w}_k \in \mathbb{C}^{N_\mathrm{Ant} \times 1}$ pointing towards the physical azimuth angle $\phi_k$ is defined as
    \begin{equation}
        \mathbf{w}_k = \frac{1}{\sqrt{N_\mathrm{Ant}}} \left[1, e^{j \frac{2\pi}{\lambda} d \sin(\phi_k)}, \dots, e^{j \frac{2\pi}{\lambda} (N_\mathrm{Ant}-1) d \sin(\phi_k)}\right]^\mathsf{T},
    \end{equation}
    where $\lambda$ is the carrier wavelength and $d$ is the antenna inter-element spacing, typically set to $d = \lambda/2$.
    
    Due to the severe free-space path loss and high susceptibility to blockages at mmWave frequencies, the channel exhibits high spatial sparsity, meaning that only a limited number of propagation paths contribute significantly to the received signal. Although channel evolution is often approximated using simplified models \cite{WSSUS, Jakes, markov}, realistic mmWave propagation can be strongly affected by user mobility, sudden blockages, and environmental changes. Therefore, we adopt the following narrowband downlink geometric channel model $\mathbf{h}[t] \in \mathbb{C}^{N_\mathrm{Ant} \times 1}$ to characterize the instantaneous propagation structure at time step $t$:
    \begin{equation}
        \mathbf{h}[t] = \sqrt{\frac{N_\mathrm{Ant}}{L[t]}} \sum_{l=1}^{L[t]} \alpha_l[t] \mathbf{a}(\theta_l[t]),
    \end{equation}
    where $L[t]$ is the number of effective multipath components, typically comprising a dominant line-of-sight (LoS) path and a few non-line-of-sight (NLoS) paths. For the $l$-th path, $\alpha_{l}[t]$ denotes the complex channel gain capturing both attenuation and phase shift, $\theta_{l}[t]$ is the angle of departure (AoD) from the BS, and $\mathbf{a}(\theta_{l}[t])$ represents the array response vector. In this sense, the temporal variation is implicitly reflected by the time-dependent parameters $L[t]$, $\alpha_l[t]$, and $\theta_l[t]$.
    
    Assuming that the BS transmits the information symbol $s[t]$ using the $k$-th beam with transmit power $P_{\mathrm{tx}}$, the received signal at the UE is expressed as
    \begin{equation}
        y_{k}[t] = \sqrt{P_{\mathrm{tx}}} \mathbf{h}[t]^\mathsf{H} \mathbf{w}_k s[t] + n[t],
    \end{equation}
    where $n[t] \sim \mathcal{CN}(0, \sigma^2)$ is the additive white Gaussian noise. The transmitted symbol $s[t]$ satisfies the normalized power constraint $\mathbb{E}[|s[t]|^2] = 1$. The optimal beam alignment task seeks the beam index $k^\star[t]$ that maximizes the reference signal received power (RSRP), i.e.,
    \begin{equation}
        \label{eq:optimal_beam}
        k^\star[t] = \underset{k \in \{1, \dots, K\}}{\arg\max} \, |\mathbf{h}[t]^\mathsf{H} \mathbf{w}_k|^2.
    \end{equation}
    
    \subsection{Multimodal Sensing Observation Space}
    To provide rich environmental and state priors and minimize the reliance on exhaustive beam training, we take advantage of readily available multimodal sensing data. Specifically, the BS collects time-synchronized OOB sensing data, including image, radar, and LiDAR measurements, while the UE acquires its GNSS position readings and feeds them back to the BS. Concurrently, the communication system captures in-band RF measurements to provide PHY context. At time step $t$, the raw multimodal observation is denoted by $\mathbf{O}[t] = \{\mathbf{X}_{m}[t]\}_{m \in \mathcal{M}}$, where $\mathcal{M} = \{\mathrm{Image}, \mathrm{Radar}, \mathrm{LiDAR}, \mathrm{GPS}, \mathrm{RF}\}$. Specifically, it includes:
    \begin{enumerate}
        \item $\mathbf{X}_\mathrm{Image}[t] \in \mathbb{R}^{3 \times H_\mathrm{Image} \times W_\mathrm{Image}}$: an RGB image captured by a camera at the BS side, providing visual information about the surrounding environment.
        
        \item $\mathbf{X}_\mathrm{Radar}[t] \in \mathbb{C}^{N_\mathrm{rx} \times N_\mathrm{chirp} \times N_\mathrm{adc}}$: the raw complex intermediate-frequency (IF) radar measurements collected by the BS-side frequency-modulated continuous-wave (FMCW) radar, where $N_\mathrm{rx}$, $N_\mathrm{chirp}$, and $N_\mathrm{adc}$ denote the numbers of receive antennas, chirps, and analog-to-digital converter (ADC) samples, respectively.
        
        \item $\mathbf{X}_\mathrm{LiDAR}[t] \in \mathbb{R}^{3 \times N_\mathrm{pts}}$: the three-dimensional (3D) point cloud collected by the BS-side LiDAR at time $t$, containing the Cartesian coordinates of $N_\mathrm{pts}$ reflected points and describing the geometry of surrounding objects and potential blockages.
        
        \item $\mathbf{x}_\mathrm{GPS}[t] = [x_\mathrm{lon}[t], x_\mathrm{lat}[t]]^\mathsf{T} \in \mathbb{R}^2$: the GNSS longitude and latitude readings acquired at the UE side and fed back to the BS, providing coarse location information of the UE in a noisy environment.
        
        \item $\mathbf{x}_\mathrm{RF}[t] \in \mathbb{R}^K$: the RF beam-level RSRP scan vector obtained from in-band measurements over previous transmission intervals, serving as historical communication feedback.
    \end{enumerate}

    \subsection{Problem Formulation}
    Let $T_{\mathrm{Hist}}$ denote the number of observed historical frames and $T_{\mathrm{Pred}}$ denote the number of future prediction frames. We consider a discrete-time multimodal sequence of total length
    \begin{align}
        T = T_{\mathrm{Hist}} + T_{\mathrm{Pred}},
    \end{align}
    sampled with a fixed time interval. The historical observation window corresponds to time steps $t \in \{1,\dots,T_{\mathrm{Hist}}\}$, while the future prediction window corresponds to time steps $\tau \in \{T_{\mathrm{Hist}}+1,\dots,T\}$. At the current time step $t$, the available historical context is defined as $\mathcal{O}[t]=\{\mathbf{O}[t-T_{\mathrm{Hist}}+1],\ldots,\mathbf{O}[t]\}$.
    
    The goal is to predict the future UE trajectory, the future optimal beam, and the future received signal strength over the next $T_{\mathrm{Pred}}$ steps.\footnote{We include RSSI prediction as a compact link-strength prediction task. Prior vision-aided studies have used received power and RSSI as a meaningful communication target, since it is relevant to downstream functions such as power control, link budget estimation, and network planning \cite{readradio,vision_CE}.} Specifically, for each future step $\tau \in \{1,\dots,T_{\mathrm{Pred}}\}$, $\mathbf{Y}_{\mathrm{Loc}}[\tau] \in \mathbb{R}^{2}$ denotes the UE location at time $T_{\mathrm{Hist}}+\tau$, $\mathbf{Y}_{\mathrm{Beam}}[\tau,:] \in \{0,1\}^{K}$ denotes the one-hot beam label at time $T_{\mathrm{Hist}}+\tau$, whose nonzero entry indicates the optimal beam among the $K$ candidate beams, and $\mathbf{y}_{\mathrm{RSSI}}[\tau] \in \mathbb{R}$ denotes the received signal strength at time $T_{\mathrm{Hist}}+\tau$. Accordingly, the stacked future targets are denoted by $\mathbf{Y}_{\mathrm{Loc}} \in \mathbb{R}^{T_{\mathrm{Pred}}\times 2}$, $\mathbf{Y}_{\mathrm{Beam}} \in \{0,1\}^{T_{\mathrm{Pred}}\times K}$, and $\mathbf{y}_{\mathrm{RSSI}} \in \mathbb{R}^{T_{\mathrm{Pred}}}$.
    
    Accordingly, the predictive modeling objective can be written as
    \begin{align}
    \max_{\Theta}\;\mathbb{P}\big(\mathbf{Y}_{\mathrm{Loc}},\mathbf{Y}_{\mathrm{Beam}},\mathbf{y}_{\mathrm{RSSI}}\mid\mathcal{O}[t];\Theta\big),
    \end{align}
    where $\Theta$ denotes the collection of all trainable parameters of the predictive model.
    
\section{JEPA-MSAC Architecture}
\label{sec:architecture}
\begin{figure*}[t]
        \centering
        \captionsetup{font=footnotesize}
        \includegraphics[width=0.99\textwidth]{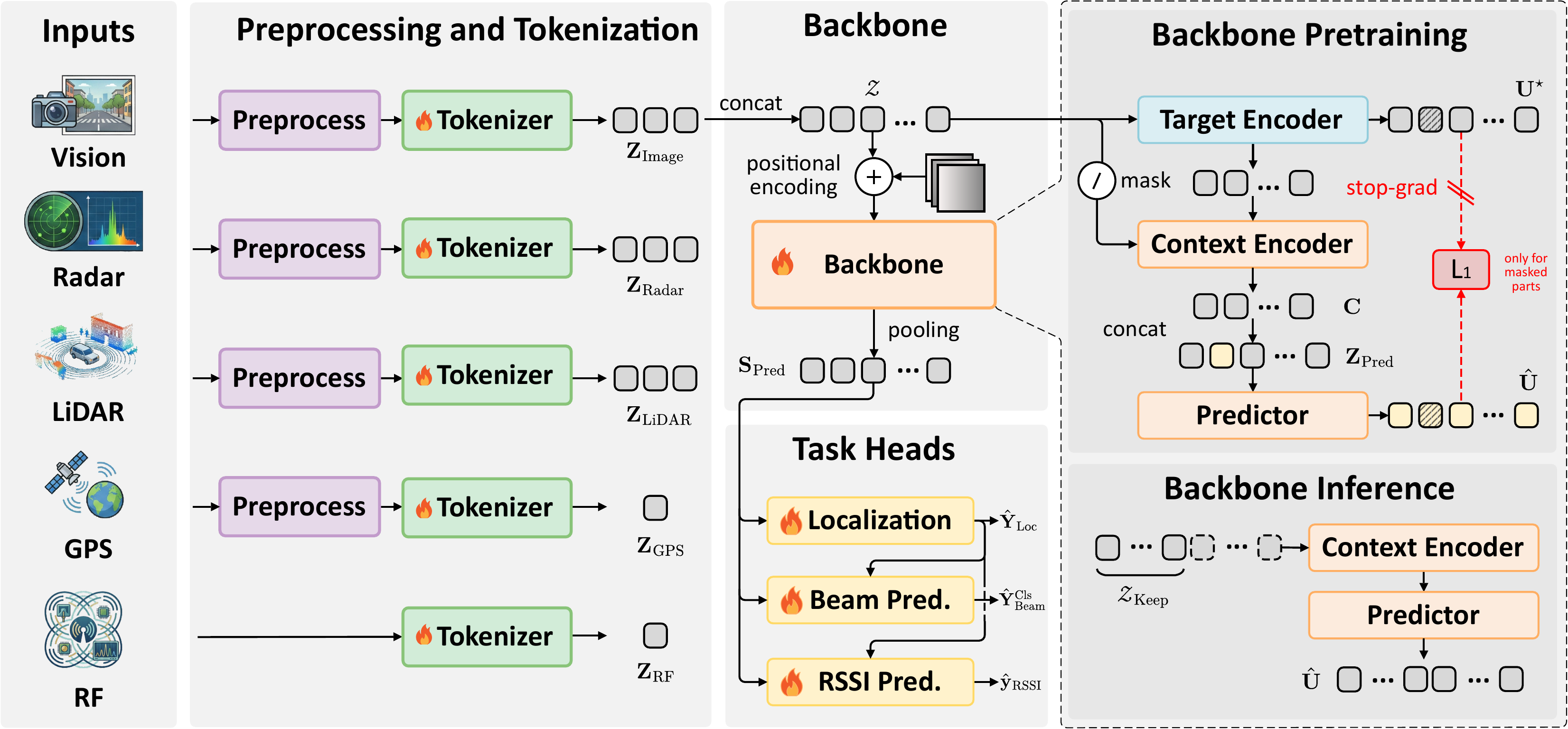}
        \caption{Overall framework of the proposed JEPA-MSAC. JEPA-MSAC tokenizes multimodal observations, pretrains a JEPA backbone, and adapts frozen features to localization, beam prediction, and RSSI prediction. Modules marked with a spark icon are learnable.}
        \label{fig:framework}
    \end{figure*}

    As shown in Fig.~\ref{fig:framework}, JEPA-MSAC first pretrains a JEPA backbone on multimodal inputs and then freezes it to generate future features for downstream PHY-tasks.

    \subsection{Multimodal Data Preprocessing}
    Since raw multimodal data vary in dimensionality and physical meaning, JEPA-MSAC applies modality-specific preprocessing to convert these inputs into structured formats suitable for tokenization.

    \subsubsection{Vision Preprocessing}
    For the raw high-resolution image $\mathbf{X}_{\mathrm{Image}}[t]$, we first crop and resize it to a standard DL resolution of $H_V \times W_V$ using bilinear interpolation. Then, each RGB channel is normalized using predefined channel-wise mean and standard deviation values:
    \begin{align}
        \mathbf{V}[t]
        =
        \left(\mathrm{Resize}(\mathbf{X}_{\mathrm{Image}}[t]) \ominus \boldsymbol{\mu}_V\right)
        \oslash
        \boldsymbol{\sigma}_V,
    \end{align}
    where $\boldsymbol{\mu}_V \in \mathbb{R}^{3}$ and $\boldsymbol{\sigma}_V \in \mathbb{R}^{3}$ denote the channel-wise mean and standard deviation vectors, respectively. The subtraction and division are applied channel-wise with broadcasting over the spatial dimensions reduces illumination bias unrelated to geometry and occlusion, generating a normalized visual tensor $\mathbf{V}[t] \in \mathbb{R}^{3 \times H_V \times W_V}$ that captures semantic features.

    \subsubsection{FMCW Radar Range-Angle Mapping}
    The raw radar signal is converted into a range-angle map via standard processing. Let $\mathbf{X}_{\mathrm{Radar}}^{(c)}[t] \in \mathbb{C}^{N_{\mathrm{rx}} \times N_{\mathrm{adc}}}$ denote the raw radar data slice corresponding to the $c$-th chirp at time $t$. First, we apply a discrete Fourier transform (DFT) along the fast-time dimension to extract range bins, and subtract the antenna mean to remove static clutter. A second DFT is then applied across the spatial dimension to resolve the angular bins. Averaging the magnitudes over $N_\mathrm{chirp}$ chirps yields the final map $\mathbf{R}[t] \in \mathbb{R}^{A_\mathrm{Radar} \times R_\mathrm{Radar}}$, where $A_\mathrm{Radar}$ and $R_\mathrm{Radar}$ are the angular and range bin counts, respectively. This mapping is mathematically expressed as:
    \begin{align}
        \mathbf{R}(i,j)[t] = \frac{1}{N_\mathrm{chirp}} \sum_{c=1}^{N_\mathrm{chirp}} \left| \mathbf{w}_i^\mathsf{H} \mathbf{X}_{\mathrm{Radar}}^{(c)}[t] \mathbf{w}_j \right|,
    \end{align}
    where $\mathbf{w}_i$ and $\mathbf{w}_j$ denote the DFT bases associated with the $i$-th angular bin and the $j$-th range bin, respectively.

    \subsubsection{3D LiDAR Point Cloud Depth Projection}
    The raw 3D LiDAR point cloud $\mathbf{X}_{\mathrm{LiDAR}}[t]$ is unordered and irregular, making direct processing computationally inefficient. Therefore, we project it onto a structured 2D depth image. Specifically, each 3D point is first converted from Cartesian coordinates to spherical coordinates consisting of radial distance $r_{\mathrm{LiDAR}}$, yaw angle $\phi_{\mathrm{yaw}}$, and pitch angle $\phi_{\mathrm{pitch}}$. Given the upward and downward viewing angles $(f_{\mathrm{up}}, f_{\mathrm{down}})$ and the maximum sensing range $d_{\max}$, the point is projected to a discrete image plane of size $H_{\mathrm{LiDAR}} \times W_{\mathrm{LiDAR}}$ as
    
    \begin{align}
        u_{\mathrm{LiDAR}} &= \left\lfloor \frac{\phi_\mathrm{yaw} + \pi}{2\pi} \times W_\mathrm{LiDAR} \right\rfloor,\\ 
        v_{\mathrm{LiDAR}} &= \left\lfloor \left(1 - \frac{\phi_\mathrm{pitch} - f_\mathrm{down}}{f_\mathrm{up} - f_\mathrm{down}}\right) \times H_\mathrm{LiDAR} \right\rfloor.
    \end{align}

    Through this depth projection, each pixel $(u_{\mathrm{LiDAR}}, v_{\mathrm{LiDAR}})$ retains the nearest normalized distance $r_{\mathrm{LiDAR}}/d_\mathrm{max}$, generating a depth matrix $\mathbf{L}[t] \in \mathbb{R}^{1 \times H_\mathrm{LiDAR} \times W_\mathrm{LiDAR}}$, perfectly preserving the physical occlusion contour of LoS propagation.

    \subsubsection{GPS Equirectangular Projection}
    For the raw GPS latitude-longitude observation $\mathbf{x}_{\mathrm{GPS}}[t]$, we transform GPS measurements into a local Cartesian coordinate system centered at the BS reference position $\mathbf{x}_{\mathrm{BS}}=[x_{\mathrm{lon}}^{\mathrm{BS}}, x_{\mathrm{lat}}^{\mathrm{BS}}]^\mathsf{T}$ via equirectangular projection:
     \begin{align}
        \mathbf{p}[t] = \frac{\pi R_E}{180} \mathbf{D} (\mathbf{x}_\mathrm{GPS}[t] - \mathbf{x}_\mathrm{BS}),
     \end{align}
     where $\mathbf{D} = \mathrm{diag}(\cos(x_\mathrm{lat}^\mathrm{BS} \cdot \frac{\pi}{180}), 1)$, and $R_E$ is the average Earth radius. This yields a local relative coordinate sequence $\mathbf{p}[t] \in \mathbb{R}^{2}$ suitable for downstream motion modeling.

    \subsection{Multimodal Tokenization}
    We use $t'$ to denote a generic time-step index within the input sequence of the tokenizer. The multimodal observation at time step $t'$ is defined as $\mathcal{O}[t'] = \{\mathbf{V}[t'], \mathbf{R}[t'], \mathbf{L}[t'], \mathbf{p}[t'], \mathbf{x}_{\mathrm{RF}}[t']\}.$ The valid range of $t'$ depends on the stage of the pipeline. During JEPA pretraining, tokenization is applied to the full multimodal sequence with $t' \in \{1,\dots,T\}$. During downstream task-head training and inference, the frozen backbone only receives the observed history as input, and thus $t' \in \{1,\dots,T_{\mathrm{Hist}}\}$.

    Since the above modalities still exhibit significant heterogeneity in spatial dimension, sampling structure, and information density, direct fusion in the original feature space would produce a high-dimensional and poorly organized representation, making learning inefficient and disrupting spatiotemporal alignment. Therefore, JEPA-MSAC designs a set of modality-specific tokenizers to map high-dimensional observations into a common token embedding space with embedding dimension of $D$:
    \begin{enumerate}
        \item Vision tokenizer: An EfficientNet pretrained on ImageNet extracts features from image $\mathbf{V}[t']$ \cite{ImageNet,efficientnet}, which are further reduced by adaptive average pooling and a $1 \times 1$ projection. These are then flattened into $P_{\mathrm{Image}}$ visual tokens, i.e., $\mathbf{Z}_\mathrm{Image}[t'] \in \mathbb{R}^{P_\mathrm{Image} \times D}$.

        \item Radar/LiDAR tokenizer: For the range-angle map generated by radar and the projected map generated by LiDAR, since both reflect spatial geometric features, we design a lightweight 2D convolutional neural network (CNN) architecture with a shared topology. After convolution, pooling, and feature mapping, the feature maps are flattened into $P_{\mathrm{Radar}}$ radar tokens and $P_{\mathrm{LiDAR}}$ LiDAR tokens, respectively. The resulting representations are $\mathbf{Z}_\mathrm{Radar}[t'] \in \mathbb{R}^{P_\mathrm{Radar} \times D}$ and $\mathbf{Z}_\mathrm{LiDAR}[t'] \in \mathbb{R}^{P_\mathrm{LiDAR} \times D}$.

        \item State tokenizer: For the low-dimensional coordinate sequence $\mathbf{p}[t'] \in \mathbb{R}^{2}$ and beam power vector $\mathbf{x}_{\mathrm{RF}}[t'] \in \mathbb{R}^{K}$, we apply a linear projection followed by layer normalization. This maps each modality at each frame into one global token, denoted as $\mathbf{Z}_\mathrm{GPS}[t'] \in \mathbb{R}^{P_\mathrm{GPS} \times D}$ and $\mathbf{Z}_\mathrm{RF}[t'] \in \mathbb{R}^{P_\mathrm{RF} \times D}$, where $P_{\mathrm{GPS}}$ and $P_{\mathrm{RF}}$ denote the numbers of GPS and RF tokens, respectively. 
    \end{enumerate}

    Through tokenization, each modality $m$ at time step $t'$ is mapped to a token tensor $\mathbf{Z}_m[t'] \in \mathbb{R}^{P_m \times D}$, where token count $P_m$ varies across modalities. We then concatenate these tensors across all modalities and time steps to form a unified sequence $\mathcal{Z}=\{\mathbf{Z}_m[t']\mid m\in\mathcal{M},\, t'=1,\dots,T\}$. This design balances representation capacity and sequence efficiency. High-dimensional spatial modalities (such as vision, radar, and LiDAR) use multiple tokens to preserve structured scene information. On the contrary, low-dimensional state modalities (such as position and RF) use a single token per frame. Ultimately, despite having different token counts, all tokens are represented in a common $D$-dimensional token space, so that they can be concatenated and jointly processed by the shared transformer backbone.

    \subsection{Factorized Positional Embedding}
    Since the input sequence of JEPA-MSAC is a long sequence composed of multiple modalities, time steps, and spatial patches, traditional absolute positional encoding cannot accurately locate the physical semantics of a particular token. Therefore, we propose a decompositional positional encoding mechanism.

    For a token $\mathbf{Z}_{m,p}[t']$ belonging to modality $m$, time step $t$, and intra-frame token index $p$, its final embedding is defined as
    \begin{align}
        \tilde{\mathbf{Z}}_{m,p}[t'] = \mathbf{Z}_{m,p}[t'] + \mathbf{E}_\mathrm{t}(t') + \mathbf{E}_\mathrm{m}(m) + \mathbf{E}_\mathrm{p}(p), \label{FPE}
    \end{align}
    where $\mathbf{E}_\mathrm{t} \in \mathbb{R}^{T \times D}$, $\mathbf{E}_\mathrm{m} \in \mathbb{R}^{|\mathcal{M}| \times D}$, and $\mathbf{E}_\mathrm{p} \in \mathbb{R}^{\max(P_m) \times D}$ are learnable temporal, modality, and intra-frame positional embedding tables, respectively. Specifically, $\mathbf{E}_\mathrm{t}$ encodes the absolute time step, $\mathbf{E}_\mathrm{m}$ identifies the modality type, and $\mathbf{E}_\mathrm{p}$ captures the intra-frame token position within each modality. Through this decoupling, the model can accurately capture complex relationships such as cross-modal temporal synchronization and intra-frame spatial cross-modal fusion in the self-attention mechanism without additional computational overhead.

    \subsection{Temporal Block-Masked JEPA Pretraining}
    \label{sec:3d}

    To equip JEPA-MSAC with predictive representations of environment dynamics, we pretrain the multimodal transformer backbone. Following the standard JEPA paradigm, we utilize an online context encoder, a momentum-updated target encoder, and a predictor for latent representation matching. This pretraining design is driven by two key considerations:

    \textbf{Latent prediction over data reconstruction:} Latent prediction is preferred to raw-signal reconstruction because downstream PHY tasks depend mainly on geometric and structural information rather than low-level details.
   
    \textbf{Temporal block masking:} Temporal block masking prevents trivial interpolation from adjacent frames and encourages the model to learn coherent temporal and cross-modal dynamics.
    
    Formally, let $\tilde{\mathcal{Z}}=\{\tilde{\mathbf{z}}_i\}_{i=1}^{I}\in\mathbb{R}^{I\times D}$ denote the multimodal token sequence after tokenization and factorized positional encoding, where $I=T\sum_{m\in\mathcal{M}}P_m$ is the total number of tokens in the unified sequence and $i$ indexes the token position. To construct the predictive task, we independently sample one contiguous temporal segment for each modality. For modality $m$, a temporal segment of length $T_{\mathrm{Mask}}=\lfloor \rho T \rfloor$ is selected, where $\rho\in(0,1)$ is the masking ratio. All tokens of modality $m$ within the selected frames are masked together. We use $\mathcal{I}_{\mathrm{Mask}}$ and $\mathcal{I}_{\mathrm{Keep}}$ to denote the index sets of masked and visible tokens, respectively, with $\mathcal{I}_{\mathrm{Mask}} \cup \mathcal{I}_{\mathrm{Keep}}=\{1,\dots,I\}$ and $\mathcal{I}_{\mathrm{Mask}} \cap \mathcal{I}_{\mathrm{Keep}}=\emptyset$. Accordingly, the masked-token subset and visible-token subset are denoted by $\mathcal{Z}_{\mathrm{Mask}}=\{\tilde{\mathbf{z}}_i \mid i\in\mathcal{I}_{\mathrm{Mask}}\}$ and $\mathcal{Z}_{\mathrm{Keep}}=\{\tilde{\mathbf{z}}_i \mid i\in\mathcal{I}_{\mathrm{Keep}}\}$, respectively.

    The JEPA pretraining consists of three components: a context encoder $f_{\theta}$, a target encoder $f_{\bar{\theta}}$, and a predictor $g_{\phi}$. The target encoder has the same architecture as the context encoder, and is subsequently updated as an exponential moving average (EMA) of the context encoder. Given the visible token sequence $\mathcal{Z}_{\mathrm{Keep}}$, the context encoder produces the contextual representation
    \begin{align}
        \mathbf{C}=f_{\theta}(\mathcal{Z}_{\mathrm{Keep}}) \in \mathbb{R}^{|\mathcal{I}_\mathrm{Keep}|\times D}.
        \label{eq:context}
    \end{align}
    To predict the masked region, we construct a full predictor input sequence $\mathbf{C}_{\mathrm{Pred}}\in\mathbb{R}^{I\times D}$ by placing the contextual representation at visible positions and a shared learnable mask token $\mathbf{z}_{\mathrm{mask}}$ at masked positions:
    \begin{align}
        \mathbf{C}_{\mathrm{Pred}}[i]=
        \begin{cases}
            \mathbf{C}[i], & i\in\mathcal{I}_{\mathrm{Keep}},\\
            \mathbf{z}_{\mathrm{mask}}, & i\in\mathcal{I}_{\mathrm{Mask}}.
        \end{cases}
        \label{eq:zpred}
    \end{align}
    
    The predictor then outputs the full latent sequence
    \begin{align}
        \hat{\mathbf{U}} = g_{\phi}(\mathbf{C}_{\mathrm{Pred}})\in\mathbb{R}^{I\times D}.
        \label{eq:predict}
    \end{align}
    Intuitively, the predictor answers the following question: \emph{given what has been observed, what should the latent representation of the missing part look like?} 
    
    In parallel, the target encoder processes the complete sequence and produces the target latent sequence
    \begin{align}
        \mathbf{U}^{\star} = f_{\bar{\theta}}(\tilde{\mathcal{Z}})\in\mathbb{R}^{I\times D}.
    \end{align}
    
    The JEPA objective trains the context encoder $f_{\theta}(\cdot)$ and predictor $g_{\phi}(\cdot)$ by aligning the masked predictions with the corresponding target latent representations:
    \begin{align}
        \mathcal{L}_{\mathrm{JEPA}}
        =
        \sum_{i\in\mathcal{I}_{\mathrm{Mask}}}
        \mathrm{Smooth}_{\ell_1}\!\left(
        \hat{\mathbf{U}}[i],
        \mathbf{U}^{\star}[i]
        \right),
    \end{align}
    where $\mathrm{Smooth}_{\ell_1}(\cdot)$ operates element-wise. For any scalar prediction $\hat{x}$ and target $x$, it is defined as:
    \begin{align}
        \mathrm{Smooth}_{\ell_1}(\hat{x},x)=
        \begin{cases}
            0.5(\hat{x}-x)^2, & \text{if } |\hat{x}-x| < 1,\\
            |\hat{x}-x| - 0.5, & \text{otherwise}.
        \end{cases}
    \end{align}
        
    Meanwhile, the target encoder $f_{\bar{\theta}}$ is updated via EMA and serves as a slowly evolving teacher that provides stable latent prediction targets, i.e.,
    \begin{align}
        \bar{\theta} \leftarrow \beta \bar{\theta} + (1-\beta)\theta.
    \end{align}
    where $\beta$ is the momentum coefficient. This momentum teacher plays a dual role. First, it stabilizes the target representation, which is critical for learning a predictive latent space without collapse. Second, it encourages the online encoder to align with a temporally smoother and more stable representation trajectory, thereby improving training robustness.

    \subsection{Frozen-Backbone Prediction}
    After JEPA pretraining, the JEPA-MSAC backbone is frozen and reused for downstream PHY-tasks. Let $\mathcal{Z}_{\mathrm{Hist}}$ denote the tokenized historical input sequence constructed from the observed multimodal context over the history window, corresponding to time steps $t=1,\dots,T_{\mathrm{Hist}}$. In the downstream stage, future prediction is formulated as a special case of the JEPA-style latent prediction process: the token positions associated with the observed history are treated as visible positions indexed by $\mathcal{I}_{\mathrm{Keep}}$, while the token positions associated with the future target slots are treated as masked positions indexed by $\mathcal{I}_{\mathrm{Mask}}$.
    Following Sec.~\ref{sec:3d}, the observed history tokens $\mathcal{Z}_{\mathrm{Hist}}$ are first encoded into the contextual representation $\mathbf{C}$ according to \eqref{eq:context}. The full predictor input sequence $\mathbf{C}_{\mathrm{Pred}}$ is then constructed in the same manner as \eqref{eq:zpred}. The frozen predictor outputs the full latent sequence as in \eqref{eq:predict}. By selecting the predicted latent representations at $\mathcal{I}_{\mathrm{Mask}}$ and aggregating the modality-specific tokens within each future frame, we obtain
    \begin{align}
        \mathbf{S}_{\mathrm{Pred}}=\mathrm{Pool}\!\left(\{\hat{\mathbf{U}}[i]\mid i\in\mathcal{I}_{\mathrm{Mask}}\}\right)\in\mathbb{R}^{T_{\mathrm{Pred}} \times D},
        \label{eq:predictive_latent_tokens}
    \end{align}
    where $\mathrm{Pool}(\cdot)$ denotes average pooling over the target-modality tokens within each future frame. $\mathbf{S}_{\mathrm{Pred}}$ serves as the predictive latent state for all downstream task-specific heads.

\begin{figure*}[t]
        \centering
        \captionsetup{font=footnotesize}
        \includegraphics[width=0.99\textwidth]{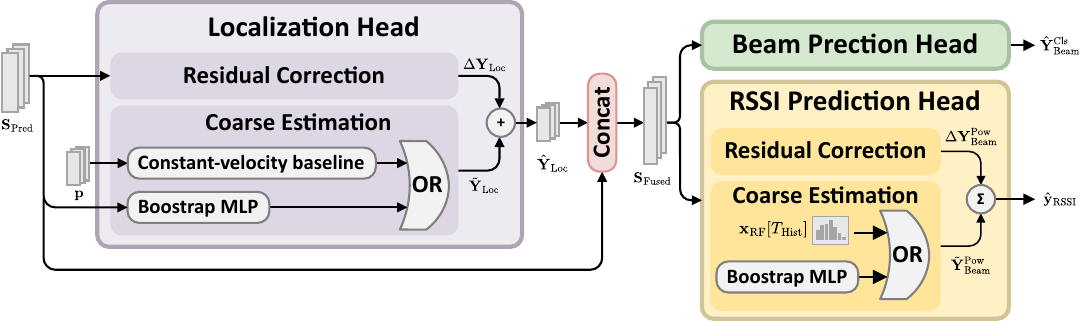}
        \caption{Overview of task-specific prediction heads on top of the predictive latent representation learned by JEPA-MSAC. The framework supports localization, beam prediction, and RSSI prediction through lightweight heads with optional localization-guided feature fusion. Residual and direct prediction modes are employed depending on the availability of historical observations.}
        \label{fig:taskhead}
    \end{figure*}

    To preserve the learned spatiotemporal structure and minimize adaptation cost, we do not finetune the backbone during downstream training. Instead, as illustrated in Fig.~\ref{fig:taskhead}, lightweight task-specific heads are trained on top of $\mathbf{S}_{\mathrm{Pred}}$ for three downstream PHY-tasks: localization, beam prediction, and RSSI prediction.

    In the standard setting, the frozen backbone receives all input modalities. We also evaluate a target-history-missing setting where historical data for the target task (e.g., location or RSSI) is unavailable. In this case, the backbone generates $\mathbf{S}_{\mathrm{Pred}}$ to bootstrap a virtual base, replacing the missing history. The task head then performs the prediction exactly as in the standard setting.

    \subsubsection{Localization/Tracking}

    For localization and tracking, the localization head predicts the future 2D trajectory $\hat{\mathbf{Y}}_{\mathrm{Loc}} \in \mathbb{R}^{T_{\mathrm{Pred}} \times 2}$ from the predictive latent sequence $\mathbf{S}_{\mathrm{Pred}}$. We adopt a coarse-estimation-plus-residual formulation. When historical UE locations are available, the coarse estimate follows a constant-velocity extrapolation
    \begin{align}
        \tilde{\mathbf{Y}}_{\mathrm{Loc}}[\tau]=\mathbf{p}[T_{\mathrm{Hist}}]+(\tau-T_{\mathrm{Hist}})\big(\mathbf{p}[T_{\mathrm{Hist}}]-\mathbf{p}[T_{\mathrm{Hist}}-1]\big),
        \label{eq:loc_cv}
    \end{align}
    for $\tau = T_{\mathrm{Hist}}+1,\dots,T$. When historical UE locations are unavailable, the coarse estimate is bootstrapped directly from the predictive latent sequence
    \begin{align}
        \tilde{\mathbf{Y}}_{\mathrm{Loc}}[\tau]=\mathrm{MLP}\!\left(\mathbf{S}_{\mathrm{Pred}}[T_{\mathrm{Pred}}]\right),
        \label{eq:loc_boot}
    \end{align}
    for $\tau = T_{\mathrm{Hist}}+1,\dots,T$, where $\mathrm{MLP}(\cdot)$ denotes the multilayer perceptron (MLP). Accordingly, the coarse trajectory estimate $\tilde{\mathbf{Y}}_{\mathrm{Loc}}$ is given by \eqref{eq:loc_cv} when location history is available, and by \eqref{eq:loc_boot} otherwise.
    
    The localization head then predicts a residual correction
    \begin{align}
        \Delta \mathbf{Y}_{\mathrm{Loc}}=h_{\mathrm{Loc}}^{\mathrm{Res}}(\mathbf{S}_{\mathrm{Pred}})\in\mathbb{R}^{T_{\mathrm{Pred}} \times 2},
        \label{eq:res_loc}
    \end{align}
    where $h_{\mathrm{Loc}}^{\mathrm{Res}}(\cdot)$ represents the residual correction layer for the localization head, and the final trajectory prediction is
    \begin{align}
        \hat{\mathbf{Y}}_{\mathrm{Loc}}=\tilde{\mathbf{Y}}_{\mathrm{Loc}}+\Delta \mathbf{Y}_{\mathrm{Loc}}.
        \label{eq:final_loc}
    \end{align}
    
    The localization head is trained with the $\ell_1$-loss
    \begin{align}
        \mathcal{L}_{\mathrm{Loc}}=\frac{1}{T_{\mathrm{Pred}}}\sum_{\tau=T_{\mathrm{Hist}}+1}^{T}\left\|\hat{\mathbf{Y}}_{\mathrm{Loc}}[\tau]-\mathbf{p}[\tau]\right\|_1.
    \end{align}

    \subsubsection{Beam Prediction}

    For beam prediction, we aim to predict the future best-beam labels. Since future beam behavior is closely related to user geometry, the predicted future coordinates from the localization branch are incorporated as auxiliary information. Specifically, the latent representation and the predicted trajectory are concatenated as $\mathbf{S}_{\mathrm{Fused}}=[\mathbf{S}_{\mathrm{Pred}};\hat{\mathbf{Y}}_{\mathrm{Loc}}]$.

    For best-beam selection, we use a lightweight classification head consisting of a projection layer, a single-layer GRU, and a decoder MLP. Given the fused representation $\mathbf{S}_{\mathrm{Fused}}$, the classification head $h_{\mathrm{Beam}}^{\mathrm{Cls}}(\cdot)$ outputs beam-selection logits over the candidate codebook:
    \begin{align}
        \hat{\mathbf{Y}}_{\mathrm{Beam}}^{\mathrm{Cls}}=h_{\mathrm{Beam}}^{\mathrm{Cls}}(\mathbf{S}_{\mathrm{Fused}})\in\mathbb{R}^{T_{\mathrm{Pred}} \times K},
    \end{align}
   where $\hat{\mathbf{Y}}_{\mathrm{Beam}}^{\mathrm{Cls}}[\tau,:]$ denotes the predicted logits over the $K$ candidate beams at the $\tau$-th future step. The corresponding ground-truth label is represented as a one-hot vector $\mathbf{Y}_{\mathrm{Beam}}[\tau,:] \in \{0,1\}^{K}$. The beam prediction head is trained as a multi-class classification task using cross-entropy loss:
    \begin{align}
        \mathcal{L}_{\mathrm{Beam}}=\frac{1}{T_{\mathrm{Pred}}}\sum_{\tau=1}^{T_{\mathrm{Pred}}}\mathrm{CE}\!\left(\hat{\mathbf{Y}}_{\mathrm{Beam}}^{\mathrm{Cls}}[\tau,:],\mathbf{Y}_{\mathrm{Beam}}[\tau,:]\right),
    \end{align}
    where $\mathrm{CE}(\cdot,\cdot)$ denotes the standard multi-class cross-entropy between the predicted beam-selection logits and the corresponding one-hot ground-truth beam label.

    \subsubsection{RSSI Prediction}

    We use a dedicated RSSI prediction head built on the same predictive latent representation. Given the fused representation $\mathbf{S}_{\mathrm{Fused}}$, the head predicts the future beam-wise power profile over the entire codebook $\hat{\mathbf{Y}}_{\mathrm{Beam}}^{\mathrm{Pow}}\in\mathbb{R}^{T_{\mathrm{Pred}} \times K}$.
    We also adopt a coarse-estimation-plus-residual formulation. The coarse beam-power estimate is defined as
    \begin{align}
        \tilde{\mathbf{Y}}_{\mathrm{Beam}}^{\mathrm{Pow}}[\tau]
        =
        \begin{cases}
            \mathbf{x}_{\mathrm{RF}}[T_{\mathrm{Hist}}],
            & \text{if RF hist. exists},\\
            \mathrm{MLP}\!\left(\mathbf{S}_{\mathrm{Pred}}[T_{\mathrm{Pred}}]\right),
            & \text{otherwise},
        \end{cases}
    \end{align}
    for all $\tau = T_{\mathrm{Hist}}+1,\dots,T$. The residual correction layer for the RSSI head $h_{\mathrm{RSSI}}^{\mathrm{Res}}(\cdot)$ then predicts a residual correction
    \begin{align}
        \Delta \mathbf{Y}_{\mathrm{Beam}}^{\mathrm{Pow}}=h_{\mathrm{RSSI}}^{\mathrm{Res}}(\mathbf{S}_{\mathrm{Fused}})\in\mathbb{R}^{T_{\mathrm{Pred}} \times K},
    \end{align}
    and the final beam-power prediction is given by
    \begin{align}
    \hat{\mathbf{Y}}_{\mathrm{Beam}}^{\mathrm{Pow}}=\tilde{\mathbf{Y}}_{\mathrm{Beam}}^{\mathrm{Pow}}+\Delta \mathbf{Y}_{\mathrm{Beam}}^{\mathrm{Pow}}.
    \end{align}
    
    The RSSI at each future step is then obtained by averaging the predicted beam-wise power values over the codebook dimension
    \begin{align}
        \hat{\mathbf{y}}_{\mathrm{RSSI}}[\tau]=\frac{1}{K}\sum_{k=1}^{K}\hat{\mathbf{Y}}_{\mathrm{Beam}}^{\mathrm{Pow}}[\tau,k].
    \end{align}
    for $\tau = T_{\mathrm{Hist}}+1,\dots,T$. The RSSI head is trained through beam-spectrum regression rather than direct scalar RSSI supervision
    \begin{align}
        \mathcal{L}_{\mathrm{RSSI}}=\frac{1}{T_{\mathrm{Pred}}}\sum_{\tau=T_{\mathrm{Hist}}+1}^{T}\mathrm{Smooth}_{\ell_1}\!\left(\hat{\mathbf{y}}_{\mathrm{RSSI}}[\tau],\mathbf{x}_{\mathrm{RF}}[\tau]
        \right).
    \end{align}
    
\section{Simulation Results}
\label{sec:simulation}
    In this section, we conduct a comprehensive evaluation of the proposed JEPA-MSAC architecture. Our experiments not only compare its predictive performance across multiple future frames but also analyze the quality of the learned latent space. Furthermore, we conduct ablation studies to verify the effectiveness of each model component.

    \subsection{Datasets and Platform}
    % 表格：实验参数配置
    \begin{table}[t]
        \centering
        \footnotesize
        \renewcommand{\arraystretch}{1.2}
        \caption{Default parameter configuration.}
        \begin{tabularx}{\linewidth}{
          >{\raggedright\arraybackslash\hsize=0.6\hsize}X 
          >{\raggedright\arraybackslash\hsize=1.3\hsize}X 
          >{\raggedright\arraybackslash\hsize=1.1\hsize}X 
        }
        \toprule
        \textbf{Category} & \textbf{Name} & \textbf{Value} \\
        \midrule
        
        \multirow{4}{=}{\textbf{Scenario Configuration}} 
        & Codebook size $K$& 64 \\
        & Total frames $T$ & 13 \\
        & Historical frames $T_\mathrm{Hist}$ & 8 \\
        & Predicted frames $T_\mathrm{Pred}$ & 5 \\
        \midrule 

        \multirow{9}{=}{\textbf{Data Preprocessing}} 
        & Vision size $H_V,W_V$ & 224,224 \\
        & Vision norm mean $\boldsymbol{\mu}_V$ & [0.485, 0.456, 0.406] \\
        & Vision norm std $\boldsymbol{\sigma}_V$ & [0.229, 0.224, 0.225] \\
        & Radar DFT size & 64 \\
        & LiDAR $H_\mathrm{LiDAR},W_\mathrm{LiDAR}$ & 64, 256\\
        & LiDAR $f_\mathrm{up},f_\mathrm{down}$ & 15$^\circ$, -15$^\circ$\\
        & LiDAR $d_\mathrm{max}$   & 100 meters\\
        & Earth's radius $R_E$ & 6371 kilometers \\
        &$P_\mathrm{Vision}$, $P_\mathrm{Radar}$, $P_\mathrm{LiDAR}$, $P_\mathrm{GPS}$, $P_\mathrm{RF}$ & 9, 16, 16, 1, 1\\
        \midrule 
        
        \multirow{6}{=}{\textbf{Stage 1: JEPA Pre-training}} 
        & Masking ratio $\rho$ & 0.5 \\
        & Epochs & 100 \\
        & Optimizer & AdamW \\
        & Learning rate & 3$\times$10$^{-4}$ \\
        & Weight decay & 0.05 \\
        & EMA momentum $\beta$& [0.996, 1.0] \\
        \midrule
        
        \multirow{3}{=}{\textbf{Stage 2: Task Head Training}} 
        & Epochs & 30 \\
        & Learning rate & $1\times10^{-4}$ \\
        & Weight decay & $1\times10^{-2}$ \\
        \bottomrule
        \end{tabularx}
        \label{tab:params}
        \end{table}
    
        \begin{table*}[t]
        \centering
        \footnotesize
        \renewcommand{\arraystretch}{1.1}
        \caption{Baseline architectures and default parameter settings.}
        
        \begin{tabular}{p{1.8cm} p{2.5cm} p{3.7cm} p{8cm}}
        \toprule
        \textbf{Task} & \textbf{Baseline} & \textbf{Input / Output} & \textbf{Architecture \& Parameters} \\
        \midrule
        
        \multirow{4}{*}{Localization}
        
        & AR
        & $(T_\mathrm{Hist},2)\rightarrow(T_\mathrm{Pred},2)$
        & Linear AR rollout with lag order $8$ and learnable weights. \\
        
        & GRU
        & $(T_\mathrm{Hist},2)\rightarrow(T_\mathrm{Pred},2)$
        & Encoder + decoder GRU $(2\rightarrow64)$ + linear head $(64\rightarrow2)$. \\
        
        & Transformer
        & $(T_\mathrm{Hist},2)\rightarrow(T_\mathrm{Pred},2)$
        & $d=64$, $n_\mathrm{head}=4$, $2$ layers, FFN $=128$. \\
        
        & Kalman
        & $(T_\mathrm{Hist},2)\rightarrow(T_\mathrm{Pred},2)$
        & constant turn rate and velocity (CTRV) state model. \\
        
        \midrule
        
        \multirow{3}{*}{Beam Prediction}
        
        & AR
        & $(T_\mathrm{Hist})\rightarrow(T_\mathrm{Pred},64)$
        & AR beam token transition with lag order $8$. \\
        
        & GRU
        & $(T_\mathrm{Hist})\rightarrow(T_\mathrm{Pred},64)$
        & Token embedding $(64\rightarrow32)$ + GRU $(32\rightarrow64)$ + linear $(64\rightarrow64)$. \\
        
        & Transformer
        & $(T_\mathrm{Hist})\rightarrow(T_\mathrm{Pred},64)$
        & Transformer decoder with embedding $64$, $n_\mathrm{head}=4$. \\
        
        & M\textsuperscript{2}BeamLLM
        & multimodal $\rightarrow(T_\mathrm{Pred},64)$
        & Refer to \cite{m2beamllm}. \\
        
        \midrule
        
        \multirow{3}{*}{RSSI Prediction}
        
        & GRU
        & $(T_\mathrm{Hist},64)\rightarrow(T_\mathrm{Pred},64)$
        & Encoder GRU $(64\rightarrow96)$ + decoder GRU $(64\rightarrow96)$ + MLP head. \\
        
        & Transformer
        & $(T_\mathrm{Hist},64)\rightarrow(T_\mathrm{Pred},64)$
        & Seq2Seq transformer: $d_\mathrm{model}=96$, $n_\mathrm{head}=4$. \\
        
        & Baseline \cite{readradio}
        & image, position $\rightarrow$ RSSI
        & MobileNetV3 visual encoder + position MLP + fusion regression head. \\
        
        \bottomrule
        \end{tabular}
        \label{tab:baseline}
        \end{table*}
        
    This work utilizes Scenario 32 of DeepSense 6G, which is a large-scale open-source dataset built upon real-world multimodal measurements \cite{deepsense6g}. The scene represents an urban street environment, where the BS is equipped with a 60 GHz mmWave phased array. We segment the continuous multimodal observation sequences, sampled at approximately 9.95 Hz, into sliding time windows of 13 frames. These window-level samples are then randomly shuffled via a configurable seed, with 70\% allocated to the training set and the remaining 30\% to the testing set. 
    The core architecture parameters of JEPA-MSAC and the hyperparameter configuration for the two-stage training are shown in Table \ref{tab:params}. In the pre-training stage, the model performs unsupervised temporal block mask JEPA training based on the full-sensory modality, and is combined with cosine annealing learning rate scheduling and momentum encoder smooth updates. In the PHY-task head training stage, we freeze the  weights of the backbone and perform supervised training only for the lightweight localization, beam, and RSSI prediction heads.

    \subsection{Baseline Schemes}

    Since each PHY-task is formulated as a sequence prediction problem, we compare our model against three common predictive baselines: an autoregressive (AR) model, a gated recurrent unit (GRU) network, and a transformer. These models capture simple step-by-step forecasting, lightweight temporal dependencies, and long-range self-attention, respectively. We also include task-specific methods for comprehensive evaluation, with all key configurations detailed in Table~\ref{tab:baseline}.
    
    For task-specific comparisons, we evaluate the following representative methods:
    \begin{itemize}
        \item \textbf{Localization:} A classical model-based Kalman filter.
        \item \textbf{Beam Prediction:} M\textsuperscript{2}BeamLLM \cite{m2beamllm}, fine-tuned with a frozen multimodal encoder, unfrozen final LLM layers, and low-rank adaptation (LoRA). We also include our RSSI prediction head, as it naturally produces $D$-dimensional outputs.
        \item \textbf{RSSI Prediction:} A vision-position baseline from \cite{readradio}, adapted by removing bounding box inputs to align with our setting while preserving the original architecture scale.
    \end{itemize}
    
    \subsection{Evaluation Metrics}
    \subsubsection{Representation Quality Metrics}

    To evaluate the pretrained representation quality, we adopt two label-free metrics: the effective rank metric ($R_{\mathrm{RankMe}}$) and the linear discriminant analysis rank metric ($R_{\mathrm{LDA}}$) \cite{RankMe, lidar}.\footnote{To avoid confusion with the LiDAR sensor, we use the abbreviation ``LDA'' for this metric instead of the ``LiDAR'' notation found in some prior literature \cite{lidar}.} Generally, higher values for both metrics indicate a superior and more diverse representation space.

    Let $\mathbf{S}=[\mathbf{s}_1^\mathsf{T};\dots;\mathbf{s}_{N_{\mathrm{Test}}}^\mathsf{T}]
    \in\mathbb{R}^{N_{\mathrm{Test}}\times D}$ be the representation matrix extracted from the frozen backbone, where $N_{\mathrm{Test}}$ is the number of test samples and $\mathbf{s}_i\in\mathbb{R}^{D}$ is the pooled representation of sample $i$.
    To compute $R_{\mathrm{RankMe}}$, we extract the singular values $\{\sigma_i\}_{i=1}^{r}$ of $\mathbf{S}$, where $r=\min(N_{\mathrm{Test}},D)$. We normalize them into a probability distribution:
    \begin{align}
        p_i=\frac{\sigma_i}{\sum_{j=1}^{r}\sigma_j}.
    \end{align}
    $R_{\mathrm{RankMe}}$ measures how evenly the representation energy distributes across latent directions. A larger value indicates a richer and less collapsed space. It is calculated as the exponential of the Shannon entropy:
    \begin{align}
        R_{\mathrm{RankMe}}=\exp\!\left(-\sum_{i=1}^{r} p_i \log p_i\right).
    \end{align}
    
    Additionally, $R_{\mathrm{LDA}}$ accounts for feature variations under data augmentations. Let $\mathbf{s}_{i,a}\in\mathbb{R}^{D}$ be the representation of the $a$-th augmented view of sample $i$. In this work, we use $A=4$ augmentations. We compute the within-augmentation covariance $\mathbf{\Sigma}_w$ and the between-sample covariance $\mathbf{\Sigma}_b$. These matrices form the transformation matrix $\mathbf{M}=\mathbf{\Sigma}_w^{-1/2}\mathbf{\Sigma}_b\mathbf{\Sigma}_w^{-1/2}$. Letting $\{\lambda_i\}_{i=1}^{D}$ be the eigenvalues of $\mathbf{M}$, we similarly normalize them into a distribution:
    \begin{align}
        q_i=\frac{\lambda_i}{\sum_{j=1}^{D}\lambda_j}.
    \end{align}
    $R_{\mathrm{LDA}}$ measures how many discriminative latent directions remain after normalizing out augmentation-induced variation. A higher value means the representations are separable across samples and stable under perturbations. Similarly, it is defined as
    \begin{align}
        R_{\mathrm{LDA}}
        =
        \exp\!\left(
        -\sum_{i=1}^{D} q_i \log q_i
        \right).
    \end{align}
    
    \subsubsection{Localization/Tracking Metrics} 
    We evaluate localization performance using two standard metrics: average displacement error (ADE) and final displacement error (FDE). Both metrics rely on the displacement error at each future step $\tau \in \{T_{\mathrm{Hist}}+1,\dots,T\}$, defined as:
    \begin{align}
        d_{\mathrm{Loc}}[\tau]=\left\|\hat{\mathbf{Y}}_{\mathrm{Loc}}[\tau]-\mathbf{p}[\tau]\right\|_{2}.
    \end{align}
    
    ADE measures the mean Euclidean error over the entire predicted trajectory:
    \begin{align}
        \mathrm{ADE}
        =
        \frac{1}{T_{\mathrm{Pred}}}
        \sum_{\tau=T_{\mathrm{Hist}}+1}^{T}
        d_{\mathrm{Loc}}[\tau].
    \end{align}
    
    Meanwhile, FDE quantifies the displacement error exactly at the final predicted step: $\mathrm{FDE}=d_{\mathrm{Loc}}[T]$.

    \subsubsection{Beam Prediction Metrics}

    Let $\mathcal{T}_{N}(\hat{\mathbf{Y}}_{\mathrm{Beam}}[\tau])$ denote the set of top-$N$ beam indices ranked by the predicted scores at time step $\tau$. The top-$N$ accuracy is defined as
    \begin{align} 
        \mathrm{ACC}_N= \frac{1}{T_{\mathrm{Pred}}} \sum_{\tau=1}^{T_{\mathrm{Pred}}} \mathbbm{1} \!_{\{ k^{\star}[\tau] \in \mathcal{T}_{N}(\hat{\mathbf{Y}}_{\mathrm{Beam}}[\tau]) \}},
    \end{align}
    where $k^{\star}[\tau]$ is the ground-truth optimal beam. We report both $\mathrm{ACC}_1$ and $\mathrm{ACC}_3$. 
    
    To evaluate beam quality from a power perspective, we compute the Layer 1 RSRP (L1-RSRP) difference. Let
    \begin{align} 
        \hat{k}[\tau] = \arg\max_{k} \hat{\mathbf{Y}}_{\mathrm{Beam}}[\tau]_{k} 
    \end{align}
    be the predicted best beam index, and let $P_k[\tau]$ be the ground-truth RSRP for beam $k$ at step $\tau$. The L1-RSRP difference is defined as
    \begin{align} 
        \Delta P[\tau] = \left| P_{\hat{k}[\tau]}[\tau] - P_{k^{\star}[\tau]}[\tau] \right|.
    \end{align}
    This metric quantifies the average power loss caused by beam selection errors.

    \subsubsection{RSSI Prediction Metrics}

    We evaluate RSSI prediction performance using root mean square error (RMSE) and mean absolute error (MAE). Both metrics depend on the absolute error at each future step $\tau \in \{T_{\mathrm{Hist}}+1,\dots,T\}$, defined as:
    \begin{align}
        d_{\mathrm{RSSI}}[\tau]=\left|\hat{\mathbf{y}}_{\mathrm{RSSI}}[\tau]-\mathbf{y}_{\mathrm{RSSI}}[\tau]\right|.
    \end{align}
    
    RMSE places greater emphasis on large prediction errors:
    \begin{align}
        \mathrm{RMSE}=\sqrt{\frac{1}{T_{\mathrm{Pred}}}\sum_{\tau=T_{\mathrm{Hist}}+1}^{T}d_{\mathrm{RSSI}}[\tau]^2}.
    \end{align}
    
    Meanwhile, MAE provides a direct measure of the average absolute deviation:
    \begin{align}
        \mathrm{MAE}=\frac{1}{T_{\mathrm{Pred}}}\sum_{\tau=T_{\mathrm{Hist}}+1}^{T}d_{\mathrm{RSSI}}[\tau].
    \end{align}
   
    \subsection{Downstream PHY-Tasks Performance}

    % 图-定位误差 CDF：不同方法
    \begin{figure}
        \centering
        \captionsetup{font=footnotesize}
        \includegraphics[width=.9\linewidth]{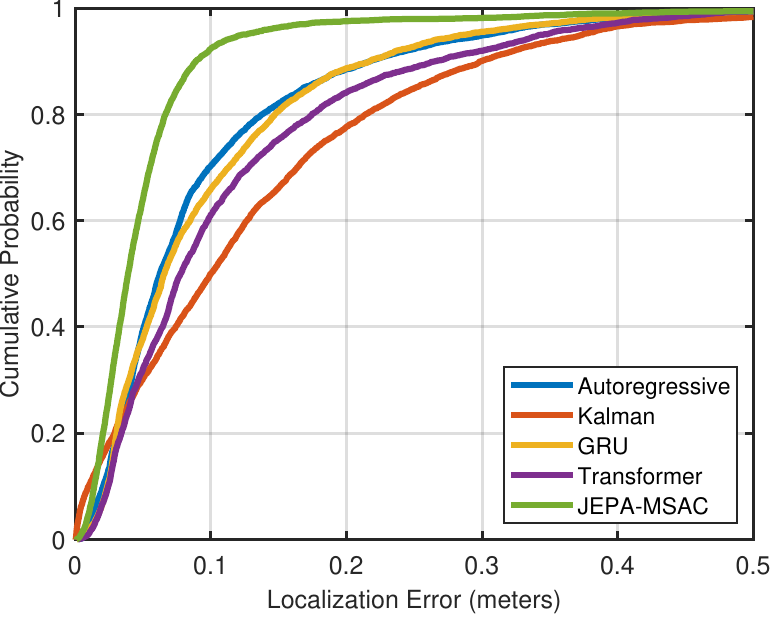}
        \caption{CDFs of the localization displacement errors for all compared methods.}
    \label{fig:loc_err_cdf}
    \end{figure}

    % 图-定位误差 vs 预测步长：不同方法
    \begin{figure}[t]
        \centering
        \captionsetup{font=footnotesize}
        \includegraphics[width=.9\linewidth]{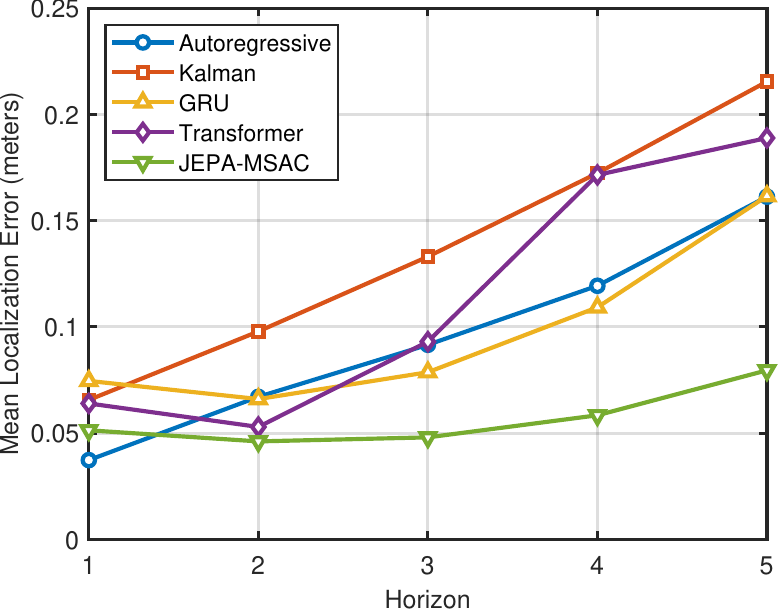}
        \caption{Mean displacement error versus prediction horizon for different prediction methods.}
        \label{fig:mean_loc_err}
    \end{figure}  

    % 图-波束预测 Top-K：不同方法
    \begin{figure}[t]
        \centering
        \captionsetup{font=footnotesize}
        
        \begin{subfigure}{0.49\linewidth}
            \centering
            \includegraphics[width=\linewidth]{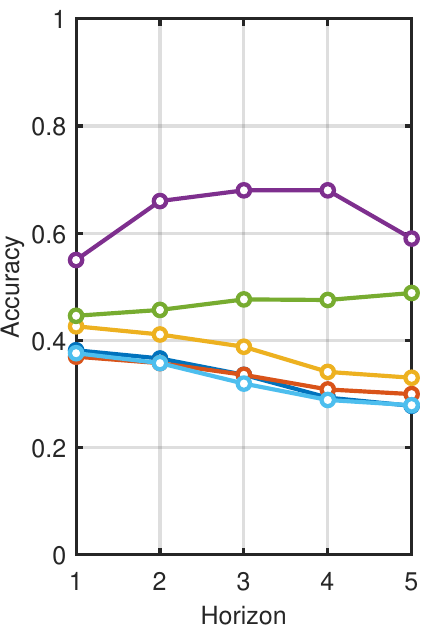}
            \caption{Top-1 accuracy.}
            \label{fig:bp_top1}
        \end{subfigure}
        \hfill
        \begin{subfigure}{0.49\linewidth}
            \centering
            \includegraphics[width=\linewidth]{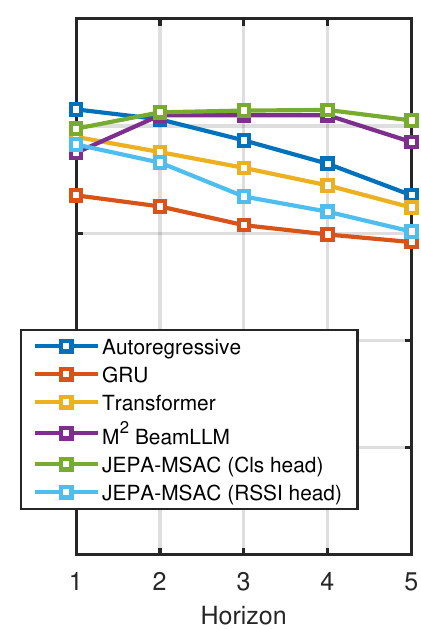}
            \caption{Top-3 accuracy.}
            \label{fig:bp_top3}
        \end{subfigure}
        
        \caption{Beam prediction accuracy versus prediction horizon under different methods.}
        \label{fig:bp_topk}
    \end{figure}
    
    % 图-波束预测 L1-RSRP：不同方法
    \begin{figure}[t]
        \centering
        \captionsetup{font=footnotesize}
        \includegraphics[width=.9\linewidth]{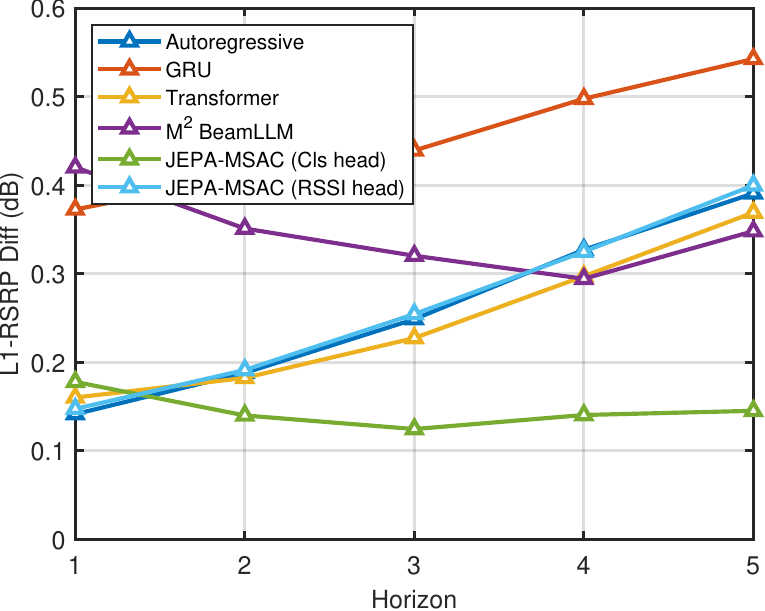}
        \caption{Mean L1-RSRP difference between the predicted and ground-truth beam power values versus prediction horizon.}
        \label{fig:l1-rsrpd}
    \end{figure} 

    \begin{figure}[t]
        \centering
        \captionsetup{font=footnotesize}
        \includegraphics[width=.9\linewidth]{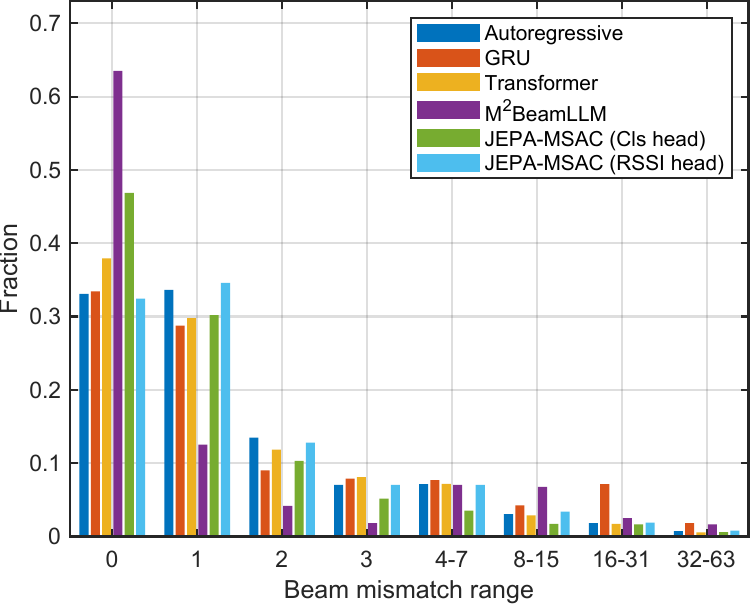}
        \caption{Distribution of the mismatch distance between the top-1 predicted beam and the ground-truth beam for different models.}
        \label{fig:beammismatch}
    \end{figure} 

    % 图-RSSI: 不同方法
    \begin{figure}[t]
        \centering
        \captionsetup{font=footnotesize}
        \begin{subfigure}{0.49\linewidth}
            \centering
            \includegraphics[width=\linewidth]{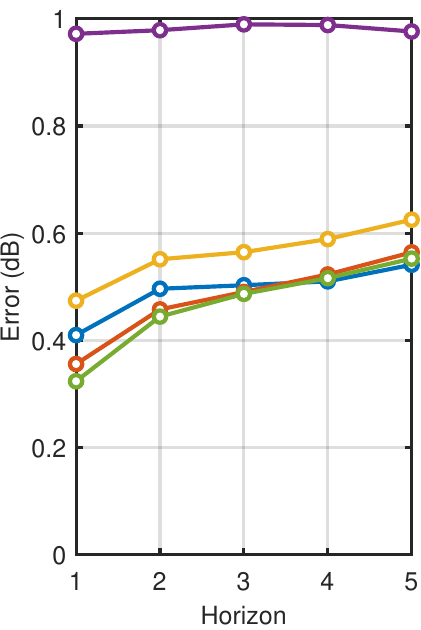}
            \caption{RMSE versus prediction horizon for different prediction methods.}
            \label{fig:rssi_rmse}
        \end{subfigure}
        \hfill
        \begin{subfigure}{0.49\linewidth}
            \centering
            \includegraphics[width=\linewidth]{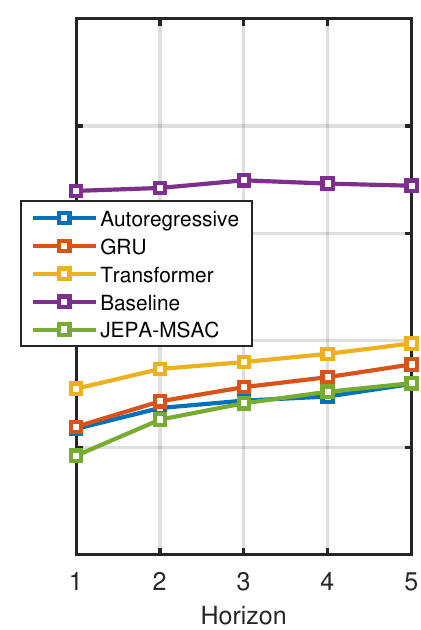}
            \caption{MAE versus prediction horizon for different prediction methods.}
            \label{fig:rssi_mae}
        \end{subfigure}
        \caption{RSSI prediction performance versus prediction horizon for different prediction methods.}
        \label{fig:rssi}
    \end{figure}

    Fig. \ref{fig:loc_err_cdf} shows the cumulative distribution functions (CDFs) of ADE, where JEPA-MSAC's steeper curve indicates superior accuracy across most test samples. Fig. \ref{fig:mean_loc_err} plots the average $d_\mathrm{Loc}$ versus the prediction horizon. While errors naturally increase over time for all methods, JEPA-MSAC consistently maintains centimeter-level accuracy and shows slower error growth than the baselines. This indicates that the learned latent representation is particularly effective for long-horizon trajectory prediction.

    Figs.~\ref{fig:bp_topk}(\subref{fig:bp_top1}) and (\subref{fig:bp_top3}) show $\mathrm{ACC}_1$ and $\mathrm{ACC}_3$ versus the prediction horizon. Overall, JEPA-MSAC consistently outperforms generic baselines. Specifically, its beam prediction head remains competitive with M\textsuperscript{2}BeamLLM and performs best among generic models, while the RSSI prediction head matches the generic baselines. Fig.~\ref{fig:l1-rsrpd} presents the mean $\Delta P$ performance. Surprisingly, despite its high accuracy, M\textsuperscript{2}BeamLLM suffers from a relatively high $\Delta P$. In contrast, the JEPA-MSAC beam prediction head achieves the lowest $\Delta P$, meaning its selected beams yield power closer to the optimal. Fig.~\ref{fig:beammismatch} explains this via the beam-mismatch distribution: JEPA-MSAC concentrates predictions in the small-mismatch region (0-3), while M\textsuperscript{2}BeamLLM and other baselines have heavier tails for larger mismatches (4-15). Thus, even when missing the exact optimal beam, JEPA-MSAC predicts nearby beams in the codebook, which better preserves received power and explains its lower $\Delta P$.
    
    Fig. \ref{fig:rssi}(\subref{fig:rssi_rmse}) shows the RMSE and MAE of RSSI prediction versus the prediction horizon. As in previous tasks, errors naturally increase over time. The performance gap between methods is smaller here. In this simple task, baseline methods also achieve good performance. However, JEPA-MSAC remains the most competitive and stable solution overall.

    \subsection{Latent Space Quality Analysis}
    % JEPA loss：不同维度
    %\begin{figure}[t]
    %    \centering
    %    \captionsetup{font=footnotesize}
    %    \includegraphics[width=\linewidth]{result_figures/fig_JEPA_loss.pdf}
    %    \caption{The JEPA loss under different embedding dimensions.}
    %    \label{fig:JEPA_loss}
    %\end{figure} 

    \begin{figure*}[htbp]
        \centering
        \captionsetup{font=footnotesize}
    
        \begin{subfigure}[t]{0.32\textwidth}
            \centering
            \includegraphics[width=\linewidth]{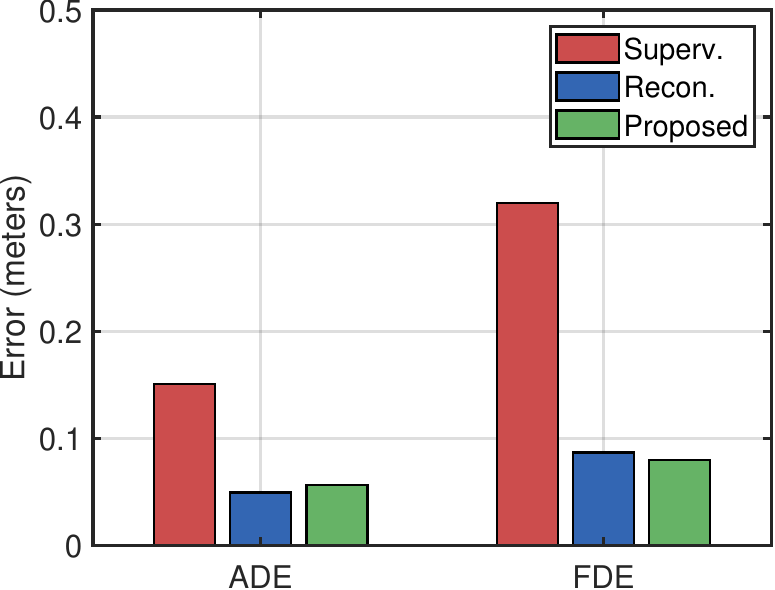}
            \caption{Localization performance.}
            \label{fig:sub_metric2}
        \end{subfigure}
        \hfill
        \begin{subfigure}[t]{0.32\textwidth}
            \centering
            \includegraphics[width=\linewidth]{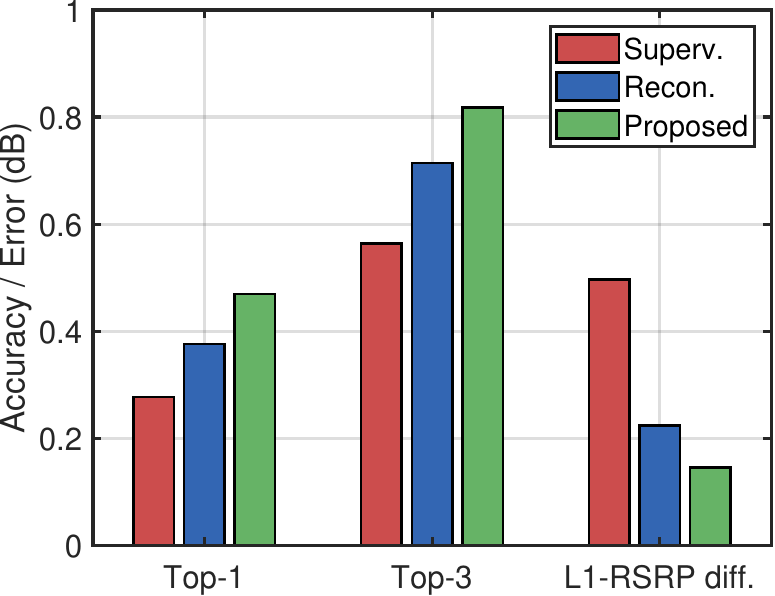}
            \caption{Beam prediction performance.}
            \label{fig:sub_metric3}
        \end{subfigure}
        \hfill
        \begin{subfigure}[t]{0.32\textwidth}
            \centering
            \includegraphics[width=\linewidth]{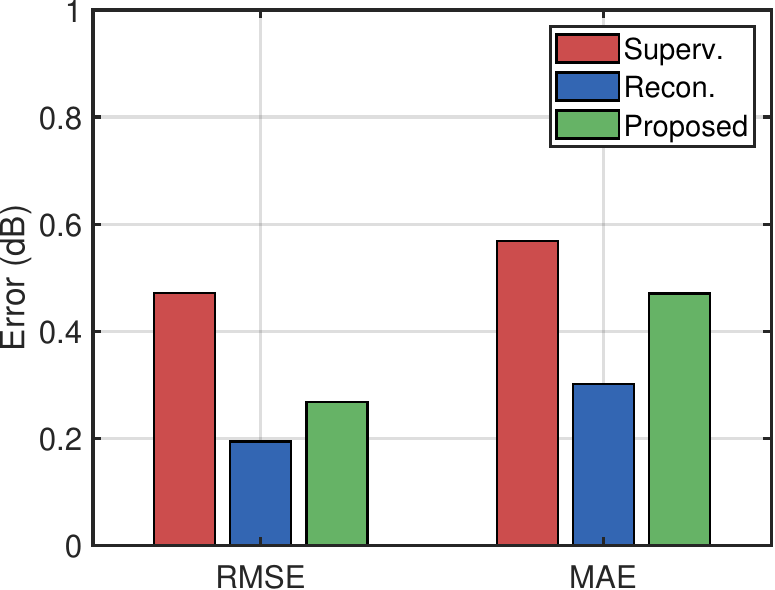}
            \caption{RSSI prediction performance.}
            \label{fig:sub_metric4}
        \end{subfigure}
    
        \caption{Performance comparison among reconstruction, supervised, and proposed methods across various metrics.}
        \label{fig:training_paradigms}
    \end{figure*}
    
    % 无监督学习表征：消融实验汇总
    \begin{table*}[t]
        \centering
        \caption{Performance comparison of models with different ablation settings across three PHY-tasks. The best result for each task is highlighted in bold.}
        \label{tab:comprehensive_ablation}
        \small
        \setlength{\tabcolsep}{4pt}
        \renewcommand{\arraystretch}{1.3} 
        
        \newcolumntype{Y}{>{\centering\arraybackslash}X}
        
        \begin{tabularx}{\linewidth}{l *{2}{Y} *{2}{Y} *{3}{Y} *{2}{Y}}
            \toprule
            \multirow{2.5}{*}{\textbf{Setting}} & \multicolumn{2}{c}{\textbf{Interpretability}} & \multicolumn{2}{c}{\textbf{Localization}} & \multicolumn{3}{c}{\textbf{Beam Prediction}} & \multicolumn{2}{c}{\textbf{RSSI Prediction}} \\
            \cmidrule(lr){2-3} \cmidrule(lr){4-5} \cmidrule(lr){6-8} \cmidrule(l){9-10}
            & $R_\mathrm{RankMe}$ $\uparrow$ & $R_\mathrm{LDA}$ $\uparrow$ & \textbf{ADE} $\downarrow$ & \textbf{FDE} $\downarrow$ & \textbf{Top-1} $\uparrow$ & \textbf{Top-3} $\uparrow$ & \textbf{L1-Diff} $\downarrow$ & \textbf{RMSE} $\downarrow$ & \textbf{MAE} $\downarrow$ \\
            \midrule
            
            % 消融：Embedding Dimensions
            $D=64$  & 7.7099 & 3.9747 & 0.0552 & 0.0927 & 0.4512 & 0.7968 & 0.1567 & 0.4703 & 0.2671 \\
            $D=128$  & \textbf{28.0365} & 7.7813 & 0.0566 & 0.0795 & 0.4692 & 0.8181 & 0.1456 & 0.4707 & 0.2674 \\
            $D=256$ & 13.3628 & 6.7065 & 0.0514 & 0.0800 & 0.4456 & 0.8023 & 0.1769 & 0.4720 & 0.2679 \\
            $D=512$ & 7.9904 & 3.1104 & \textbf{0.0502} & 0.0922 & 0.4553 & 0.7964 & 0.1672 & 0.4720 & 0.2679 \\
            \midrule
            
            % 消融：不同训练方式
            Untrained   & --  & --  & 0.0561 & 0.1060 & 0.2352 & 0.4707 & 0.9362 & 0.4695 & 0.2664 \\
            E2E       & --  & --  & 0.0548 & 0.1029 & 0.4643 & \textbf{0.8276} & 0.1520 & 0.4716 & 0.2734 \\
            \midrule
            w/o loc aux & --  & --  & -- & -- & 0.4539& 0.8018& 0.1925& 0.4715& 0.2689\\
            \midrule
            
            % 消融：模态缺失
            w/o Position    & --  & --  & 3.0416& 3.3648& -- & -- & -- & -- & -- \\
            w/o Power    & -- & --  & -- & -- & 0.4082& 0.7365& 0.1925& 0.6077& 0.3916\\
            w/o Vision    & --  & --  & 0.0712& 0.1215& 0.3941& 0.7542& 0.1787& 0.4718& 0.2677\\
            w/o Radar    & --  & --  & 0.0790& 0.0961& 0.4535& 0.8102& 0.1699& 0.4716& 0.2679\\
            w/o LiDAR    & --  & --  & 0.0750& 0.0910& 0.4651& 0.8133& 0.1447& 0.4707& 0.2672\\
            w/o V+R    & --  & --  & 0.0827& 0.1322& 0.3735& 0.7406& 0.2085& 0.4695& 0.2669\\ % done
            w/o V+L    & --  & --  & 0.0859& 0.1290& 0.3814& 0.7500& 0.1797& 0.4690& \textbf{0.2662}\\ % done
            w/o R+L    & --  & --  & 0.2771& 0.2641& 0.3885& 0.7463& 0.2544& 0.4718& 0.2677\\ % done
            \midrule
            
            % 消融：不同掩码比例
            $\rho=0.25$& 25.2351 & \textbf{8.3193} & 0.0601& 0.0832& 0.4664& 0.8144& 0.1469& 0.4692& 0.2681\\
            $\rho=0.75$& 25.6966 & 7.7024 & 0.0520& \textbf{0.0737}& \textbf{0.4699}& 0.8140& \textbf{0.1434}& 0.4716& 0.2671\\
            random mask   & 14.2474  & 7.2604  & 0.0601& 0.0860& 0.4562& 0.8119& 0.1548& \textbf{0.4666}& 0.2665\\
            checkerboard  & 18.1985  & 6.0647  & 0.0650& 0.0911& 0.4597& 0.8045& 0.1472& 0.4696& 0.2673\\
            %\midrule

            %$T_\mathrm{Hist} = 2$& --  & --  & 0.0560& 0.0834& 0.4658& 0.8111& 0.1455& 0.4701& 0.2669\\
            %$T_\mathrm{Hist} = 4$& --  & --  & 0.0570& 0.0809& 0.4676& 0.8133& 0.1449& 0.4704& 0.2671\\
            %$T_\mathrm{Hist} = 6$& --  & --  & 0.0573& 0.0807& 0.4683& 0.8162& 0.1464& 0.4705& 0.2673\\

            \bottomrule
        \end{tabularx}
    \end{table*}

    Following the overall performance comparison, we now examine how the embedding dimension affects the model. Table~\ref{tab:comprehensive_ablation} summarizes both the latent space quality metrics and the downstream task results. The quality of the latent space shows a clear non-monotonic trend. When $D$ increases from 64 to 128, both $R_\mathrm{RankMe}$ and $R_\mathrm{LDA}$ improve significantly. However, these metrics decrease when $D$ further increases to 256 and 512. Relying only on these statistical metrics is not fully convincing. We use the downstream PHY-tasks as a linear probing test to verify our analysis. As shown in the same table, different tasks react differently to dimension changes. Beam prediction performs best at $D=128$, which closely matches the representation metrics. Localization benefits from larger dimensions, while RSSI prediction hardly changes. In conclusion, an intermediate embedding dimension provides the best balance between a high-quality latent space and strong task transferability.

    \subsection{Comparison of Different Training Paradigms}

    We examine how different training paradigms affect downstream PHY-tasks. Fig.~\ref{fig:training_paradigms} compares JEPA-based predictive training (proposed), reconstruction-based self-supervised learning (SSL), and multi-task supervised learning. Multi-task learning directly optimizes downstream objectives but often forces a trade-off among different tasks. Reconstruction-based SSL retains local geometric details, which helps continuous regression tasks. In contrast, proposed JEPA-based method predicts missing latent content instead of reconstructing raw observations. This specifically encourages the model to learn temporal dynamics and cross-modal dependencies. These different mechanisms clearly impact downstream results. For localization, JEPA-MSAC and the reconstruction method achieve similar ADE. However, JEPA-MSAC shows a better FDE, demonstrating stronger long-horizon temporal modeling. For beam prediction, JEPA-MSAC shows the most significant advantage. This task relies on future propagation geometry and blockage evolution, making our predictive latent state highly suitable. Finally, reconstruction-based SSL methods achieve slightly better performance for RSSI prediction, since RSSI is a more averaged indicator of link strength.

    \subsection{Ablation Studies}
    To better understand our design choices and the effect of some parameters, we conduct a series of ablation studies. Returning to Table.~\ref{tab:comprehensive_ablation}, we further analyze the detailed settings.

    \subsubsection{Effect of Pretraining}
    We compare our unsupervised pretraining with an untrained backbone and supervised end-to-end (E2E) training. For the E2E baseline, we train a separate model for each specific task. The untrained setting performs the worst, suffering severe accuracy drops and a sharp increase in $\Delta P$ for beam prediction. Meanwhile, the task-specific E2E models achieve no significant gain over our method. While performance gaps are smaller for localization and RSSI prediction, our pretrained backbone remains the most robust overall. Furthermore, training a separate E2E model for every new task results in a much higher computational cost. In conclusion, our unsupervised approach learns strong and reusable representations. It achieves high effectiveness and efficiency across all tasks using only lightweight task-specific heads.

    \subsubsection{Effect of Localization-Guided Cascading}
    We compare the beam prediction and RSSI prediction heads with and without localization auxiliary input. Results shows that using localization input improves beam prediction, giving higher $\mathrm{ACC}_1$ and $\mathrm{ACC}_3$ and lower $\Delta P$. The effect on RSSI prediction is much smaller. Overall, localization auxiliary input is mainly helpful for beam prediction.

    \subsubsection{Effect of Modality Removal}
    
    We conduct ablation studies by removing specific input modalities. The results show that each PHY-task relies on different sensors. For localization, position data is the most critical. Removing spatial sensors like radar and LiDAR also degrades performance. For beam and RSSI prediction, RF power is the most essential input. However, beam prediction also relies on vision and other sensors, while RSSI prediction remains mostly unaffected by their removal. Overall, primary modalities dominate their specific tasks, and auxiliary sensors provide complementary information to improve robustness.

    \subsubsection{Effect of Masking}
    We evaluate mask ratios $\rho \in \{0.25, 0.5, 0.75\}$. The results show that representation metrics and PHY-task performance are not always aligned. A lower mask ratio slightly improves $R_{\mathrm{LiDAR}}$ but fails to benefit downstream tasks. Conversely, a higher mask ratio improves both localization and beam prediction. It provides lower errors, higher $\mathrm{ACC}_1$, and lower $\Delta P$. Its effect on RSSI prediction remains marginal. Overall, a higher mask ratio is more favorable for practical downstream performance. Fixing $\rho=0.5$, we also compare random, checkerboard, and temporal block masking patterns. The proposed temporal block masking achieves the highest $R_{\mathrm{RankMe}}$ and $R_{\mathrm{LDA}}$, indicating a richer and more discriminative latent space, which directly yields the best localization and beam prediction results. Random masking performs slightly better on RSSI prediction, while the checkerboard strategy shows no clear advantage. Overall, temporally contiguous masking is effective for learning predictive representations, especially for geometry-sensitive tasks.
    
    %\subsubsection{Effect of Time-Window Configuration}
    %We evaluate the model using different historical time windows: $T_{\mathrm{Hist}} \in \{2, 4, 6, 8\}$. The results suggest that a longer historical window is generally more beneficial for beam prediction, where the $\mathrm{ACC}_1$ and $\mathrm{ACC}_3$ accuracies improve gradually as more temporal context is provided. For localization, the effect is more complicated: a longer history slightly improves the long-horizon error metric, while the gain on ADE is limited. By contrast, RSSI prediction is largely insensitive to the time window, with only marginal variation across different $T_\mathrm{Hist}$. Overall, these results indicate that temporal context is helpful, but its benefit is task-dependent and tends to saturate beyond a moderate history time window.

    \begin{table}[t]
    \centering
    \footnotesize
    \renewcommand{\arraystretch}{1.12}
    \setlength{\tabcolsep}{0pt}
    \caption{Model complexity and inference latency of all compared methods, with tokenizer-only, backbone-only, and task-head-only timings reported separately.}
    \label{tab:overall_complexity}
    \begin{tabular*}{\linewidth}{@{\extracolsep{\fill}} l l c c c @{}}
    \toprule
    \textbf{Tasks} & \textbf{Method} & \multicolumn{1}{l}{\textbf{Params}} & \multicolumn{1}{l}{\textbf{FLOPS}} & \multicolumn{1}{l}{\textbf{Latency (ms)}} \\
    \midrule
    
    \multirow{5}{*}{\makecell[l]{\textbf{Tokenizer}\\\textbf{(per frame)}}}
    & Vision tokenizer & 4.17 M & 0.83 G  & 0.78 \\
    & Radar tokenizer  & 9.34 K & 7.62 M  & 0.06 \\
    & LiDAR tokenizer  & 9.34 K & 7.62 M  & 0.06 \\
    & GPS tokenizer    & 640    & 1.54 K  & 0.03 \\
    & RF tokenizer     & 8.58 K & 17.41 K & 0.03 \\
    \midrule
    
    \multirow{5}{*}{\makecell[l]{\textbf{Backbone}}}
    & $D=64$            & 0.40 M  & 37.60 M & 0.66 \\
    & $D=128$           & 1.59 M  & 0.14 G & 0.74 \\
    & $D=256$           & 6.32 M & 0.56 G & 1.22 \\
    & $D=512$           & 25.22 M & 2.19 G & 3.10 \\
    & Tar. Enc. ($D=128$) & 0.79 M  & --      & -- \\
    \midrule
    
    \multirow{8}{*}{\textbf{Loc.}}
    & AR                  & 18       & --      & 0.07 \\
    & GRU                 & 26.24 K  & 0.35 M  & 0.13 \\
    & Transformer         & 234.24 K & 3.10 M  & 1.08 \\
    & Kalman              & 1        & --      & 0.03 \\
    & Proposed ($D=64$)   & 0.46 M   & 4.51 M  & 0.11 \\
    & Proposed ($D=128$)  & 0.50 M   & 4.71 M  & 0.11 \\
    & Proposed ($D=256$)  & 0.56 M   & 5.10 M  & 0.11 \\
    & Proposed ($D=512$)  & 0.69 M   & 5.89 M  & 0.11 \\
    \midrule
    
    \multirow{8}{*}{\makecell[l]{\textbf{Beam}\\\textbf{Prediction}}}
    & AR                          & 4.45 K   & 20.48 K & 0.09 \\
    & GRU                         & 134.75 K & 1.72 M  & 0.18 \\
    & Transformer                 & 242.05 K & 3.17 M  & 1.08 \\
    & M\textsuperscript{2}BeamLLM & 50.58 M  & 29.90 G & 3.00 \\
    & Proposed ($D=64$)           & 1.76 M   & 17.64 M & 0.09 \\
    & Proposed ($D=128$)          & 1.79 M   & 17.97 M & 0.08 \\
    & Proposed ($D=256$)          & 1.86 M   & 18.62 M & 0.08 \\
    & Proposed ($D=512$)          & 1.99 M   & 19.93 M & 0.09 \\
    \midrule
    
    \multirow{8}{*}{\makecell[l]{\textbf{RSSI}\\\textbf{Prediction}}}
    & AR                        & 9        & --       & 0.09 \\
    & GRU                       & 25.79 K  & 0.35 M   & 0.16 \\
    & Transformer               & 234.05 K & 3.09 M   & 1.10 \\
    & Baseline~\cite{readradio} & 1.01 M   & 121.90 M & 0.62 \\
    & Proposed ($D=64$)         & 1.83 M   & 17.64 M  & 0.11 \\
    & Proposed ($D=128$)        & 1.89 M   & 17.97 M  & 0.11 \\
    & Proposed ($D=256$)        & 2.02 M   & 18.62 M  & 0.11 \\
    & Proposed ($D=512$)        & 2.29 M   & 19.93 M  & 0.12 \\
    \midrule
    
    \multirow{3}{*}{\makecell[l]{\textbf{Training}\\\textbf{Paradigm}}}
    & Recon. backbone      & 5.00 M  & 5.68 G  & 12.00 \\
    & Recon. decoder       & 24.71 M & --      & -- \\
    & Multitask supervised & 10.77 M & 47.85 G & 24.52 \\
    \bottomrule
    \end{tabular*}
    \end{table}

    \subsection{Complexity and Efficiency Analysis}

    Table~\ref{tab:overall_complexity} summarizes the model complexity and inference latency measured on an NVIDIA L4 GPU. For JEPA-MSAC, the computational cost is concentrated in the shared backbone, while the downstream task heads remain extremely lightweight. This design offers a clear advantage over single-task baselines like M\textsuperscript{2}BeamLLM, which require dedicated models for each task. Furthermore, JEPA-MSAC generates future predictions in a one-shot manner. This makes its inference faster than the AR decoding used by baselines. During training, JEPA-MSAC also avoids the heavy decoders required by reconstruction-based SSL and the high complexity of multitask supervised learning. Overall, JEPA-MSAC provides an highly efficient balance between reusable representation learning and low-cost task adaptation.

\section{Conclusion}
\label{sec:conclusion}
    In this paper, we proposed JEPA-MSAC, a multimodal predictive latent learning framework for sensing-assisted PHY-tasks. The framework learns a shared predictive backbone from multimodal sensing and communication observations, and supports localization, beam prediction, and RSSI prediction through low-cost task adaptation. By learning cross-modal temporal dynamics in a unified latent space, the proposed method provides a reusable representation for multiple future PHY-tasks. Extensive experiments demonstrated strong performance across all three tasks, with the proposed method outperforming most baselines in most settings. Comprehensive ablation studies further confirmed the effectiveness of the proposed design. In particular, the comparisons with multi-task supervised learning and reconstruction-based SSL showed that the main advantage comes from predictive latent state learning itself, rather than from model scale or multimodal fusion alone. Overall, this work shows that predictive latent state learning is an effective unified approach for sensing-assisted PHY-tasks.
    
    Future work will extend JEPA-MSAC from predictive representation learning to a broader communication-oriented physical AI framework. A key direction is to incorporate structured causal reasoning and closed-loop decision support on top of the learned world state, enabling the model not only to predict future PHY variables but also to support proactive communication control \cite{alpamayo}. Another important direction is to develop scalable wireless world foundation models with stronger cross-scenario transfer and simulation-to-real generalization, while covering more general settings such as different frequency bands and different numbers of users.

\balance
\bibliographystyle{IEEEtran}
\bibliography{IEEEabrv, ref}

%\ifCLASSOPTIONcaptionsoff
%  \newpage
%\fi
%\begin{thebibliography}{1}
%\end{thebibliography}
%\end{IEEEbiography}
\end{document}